\documentclass{cernyrep}
\usepackage{texnames}
\usepackage[T1]{fontenc}

\usepackage[bookmarks, colorlinks=true, linktoc=page, pdftex, linkcolor=black, citecolor=black, urlcolor=blue]{hyperref}

\usepackage{fancyhdr}
\fancyhfoffset{4 mm}
\fancypagestyle{ARTTITLE}{%
\fancyhf{} 

\lhead{\hfill CERN Accelerator School Proceedings --
{\it Intensity Limitations in Hadron Beams} --  Borovets, Bulgaria, 2025\hfill}
\lfoot{\hspace{3mm} Available online at \url{https://cas.web.cern.ch/previous-schools}}
\rfoot{\thepage\hspace*{3mm}}

}

\usepackage{varwidth}
\usepackage{xcolor}
\DeclareMathOperator{\Tr}{Tr}
\DeclareMathOperator{\sgn}{sgn}

\begin{document}
\newcommand{\mytilde}[1]{\tilde{#1}}
\newcommand{\permittivity}{\epsilon_0}

\title{Intrabeam Scattering}
 
\author{Andrzej Wolski}

\institute{University of Liverpool, and the Cockcroft Institute, UK}

\begin{abstract}
Intrabeam scattering refers to the effects of the Coulomb interaction acting
between pairs of charged particles within a bunch in an accelerator. One of the~main
consequences of intrabeam scattering is a change in the emittances of a~bunch: in
some circumstances (in particular, in hadron storage rings operating above transition),
the transverse and longitudinal emittances may grow over time without limit. This may
restrict the performance of machines for which maintaining low beam emittance is an
important requirement.  In this
note, we describe some of the models used to analyse the effects of intrabeam
scattering, and present in particular the Piwinski formulae for the emittance growth
rates.  We compare the predicted changes in emittance  with measurements in a~number of machines operating in different parameter regimes. 
\end{abstract}


\maketitle 
\thispagestyle{ARTTITLE}
 
\section{Introduction}

As a bunch of charged particles moves along an accelerator beamline, particles interact
through the~Coulomb force.  In principle, there are two ways in which the effects of the
Coulomb interaction may be modelled: either, a bunch may be represented as a continuous
charge distribution with a size and shape that evolve as the bunch moves along a
beamline; or a bunch may be represented as a collection of discrete particles that
scatter (or `collide' with each other).  The continuum model leads to the theory of
space-charge \cite{ferrario2014, franchetti2015}; the discrete particle model leads to the
effects of Touschek scattering \cite{bernardini1963} and intrabeam scattering
\cite{piwinski1974, piwinskibjorkenmtingwa2018}.

Touschek scattering and intrabeam scattering both involve transfer of energy and
momentum between the (transverse and longitudinal) degrees of freedom.  In
Touschek scattering, however, the change in longitudinal momentum in a collision
between two particles is large: in a storage ring, the change may be large enough that
the final longitudinal momentum of one or both particles lies outside the acceptance
of the ring, so that particles can be lost from the beam.  `Large angle' scattering events
leading to particle loss are infrequent compared to the rate of `small angle' scattering
events; nevertheless, the steady loss of particles through Touschek scattering (quantified
by the Touschek lifetime) is often a dominant limitation on the lifetime of beams in
low-emittance electron storage rings.

The frequent collisions leading to smaller changes in momentum (intrabeam scattering,
IBS) also have observable effects: although particles are not lost from the beam, there can
be significant changes in beam emittance.  The changes that occur depend on whether
the storage ring is operating below or above transition.  As we shall see in
Section~\ref{sec:simplegasmodel}, below transition a simple model may be developed in
which the~particles in a bunch behave as atoms or molecules in an ideal gas.  Scattering
then leads to the~particles reaching an equilibrium distribution, with the energy shared
between the different degrees of freedom.  Above transition, however, the behaviour is
very different: in that case, there is no equilibrium, but the~emittances can grow without
limit.

The timescale of changes in beam emittance from intrabeam scattering
depends on many factors, including the beam energy, the charge density within the bunch
and the storage ring optics. Over a wide range of parameter regimes, emittance growth
times are on the timescale of minutes or hours. For hadron colliders, this can limit the
luminosity lifetime and require the injection of fresh (low-emittance) bunches of particles
at regular intervals.  In electron storage rings, synchrotron radiation damping generally
dominates over emittance growth from intrabeam scattering (though particle loss from
Touschek scattering is often an important limitation on beam lifetime).  However, for the
latest generation of ultra-low emittance electron storage rings for synchrotron light
sources, the particle density can be high enough that intrabeam scattering leads to an
increase of the equilibrium emittance that would be achieved through synchrotron radiation
effects acting alone.  Intrabeam scattering has even been observed in a high-brightness
driver for a free-electron laser \cite{dimitri2020}.

The main goal of theoretical analysis of intrabeam scattering
is to determine formulae for the~emittance growth rates.  Some early analysis was done in
the late 1960's, but more thorough investigations were carried out in the early 1970's initially
motivated by the realisation that intrabeam scattering could limit the luminosity of collisions
in the SPS at CERN.  The first detailed expressions for the emittance growth rates resulting from
intrabeam scattering were derived (with some simplifications) by Anton Piwinski in 1974
\cite{piwinski1974}. Calculation of the growth rates using Piwinski's formulae in a given machine
can be computationally expensive, primarily because the growth rates are expressed in terms of
integrals that must be evaluated numerically. The integrals depend on the lattice functions,
so an accurate calculation of the growth rates requires a numerical integration to be
performed at many individual points around the~lattice. Soon after Piwinski's work, an alternative
formulation of the intrabeam scattering growth rates was developed by Bjorken and Mtingwa
\cite{piwinskibjorkenmtingwa2018}. Although the expressions for the growth rates in Bjorken and
Mtingwa's formulae take a different form from those in Piwinski's formulae, they lead to the same
results (and evaluation can still be computationally expensive). In the following years and decades,
further work aimed to improve the accuracy of the formulae by including effects neglected in the
original analysis and to simplify or approximate  the expressions to allow faster computation. This
has led to some more convenient ways to estimate the emittance growth rates from IBS in certain
regimes, for example in high-energy storage rings.

Experimental studies of intrabeam scattering are often challenging because of the difficulty
of separating different sources of emittance growth.  For example, in hadron storage rings
impedance effects and nonlinear effects can contribute to emittance growth on similar
timescales to IBS.  Nevertheless, studies have been carried out on several machines
(including proton, ion and electron storage rings) covering a wide range of parameter
regimes, with results that are generally consistent with the theories\footnote{It is often
necessary to make some assumptions for certain beam parameters when comparing
experimental results with theories of IBS, since it is not always possible to make direct
measurements of all the quantities needed for the theoretical calculations.}.  Intrabeam
scattering is now generally considered to be a well-understood phenomenon.  The fundamental
nature of the scattering process means that IBS is often a significant issue in storage rings
requiring high-intensity, low emittance beams.

In the following sections, we first present some examples of observations of intrabeam
scattering effects in different machines.  We then discuss a simple model based on treating
the particles within a~bunch as atoms or molecules in an ideal gas: although this model is
too basic to be of practical use in calculating growth rates, it does provide some insight into
some significant features of IBS.  It can help to explain, for example, why IBS leads to a beam
reaching a new equilibrium distribution in a storage ring below transition whereas above
transition the emittances can grow without limit.  After discussing the simple `ideal gas' model,
we present Piwinski's formulae for the IBS emittance growth rates.  We do not attempt to show a
full derivation, as this is rather lengthy and mathematically involved: more complete treatments
can readily be found in other places.  We also discuss some of the significant theoretical
developments aiming (for example) to provide simpler formulae for calculating the IBS growth
rates in particular regimes.  Finally, we present some examples of experimental measurements
aiming to compare the growth rates from IBS in real machines with the predictions of the
theoretical formulae.

\section{Some observations of intrabeam scattering}

Before discussing theoretical models and IBS growth rate formulae, we show some examples
of the~effects of intrabeam scattering.  For our examples, we have selected three very different
machines: two proton storage rings, one operating at 270\,GeV (the CERN SPS \cite{gareyte1984})
and the other at 400\,MeV (CELSIUS at The Svedberg Laboratory, Uppsala, Sweden
\cite{raohermansson2000}), and an electron storage ring (CESR at Cornell University, Ithaca,
New York, USA \cite{ehrlichman2013}). Some of the relevant parameters for these accelerators
are listed in Table\,\ref{tab:examplemachineparameters}.  Further examples, showing more detailed
comparisons between measurements and theoretical predictions, will be given in Section
\ref{sec:experimentalmeasurements}.

\begin{table}[b!]
\caption{Representative beam parameters in storage rings chosen to illustrate observations
of intrabeam scattering.\label{tab:examplemachineparameters}}
\begin{center}
\begin{tabular}{lcccc}
\hline
 & \textbf{SPS \cite{gareyte1984}} & \multicolumn{2}{c}{\textbf{CELSIUS \cite{raohermansson2000}}} & \textbf{CESR \cite{ehrlichman2013}} \\
 \hline
 Particle type & proton/antiproton & protons & ions & electrons/positrons\\
 Beam energy (max) & 270\,GeV &  400\,MeV & 200\,MeV/u & 2.085\,GeV \\
 Bunch population & $10^{11}$ & $10^{11}$ & $10^{10}$ &  $1.6\times 10^9 - 1.6\times 10^{11}$ \\
 Bunch length &   &  \multicolumn{2}{c}{not known}  & 10\,mm \\
 Relative energy spread &   & \multicolumn{2}{c}{$10^{-4}$ (assumed)} & $8\times 10^{-4}$ \\
 Longitudinal emittance & $1$\,eV\,s &  &  &  \\
 Transverse emittance & $20\pi$\,mm\,mrad & \multicolumn{2}{c}{$1\pi$\,mm\,mrad} & 3.4\,nm$\times$20\,pm \\
  & & & & (h$\times$v, geometric) \\
 \hline
\end{tabular}
\end{center}
\end{table}

\begin{figure}[t!]
\begin{center}
\includegraphics[width=0.300\textwidth]{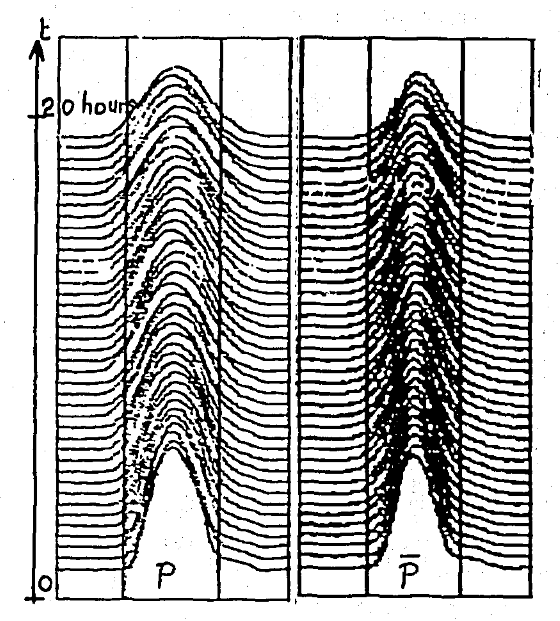}
\includegraphics[width=0.320\textwidth]{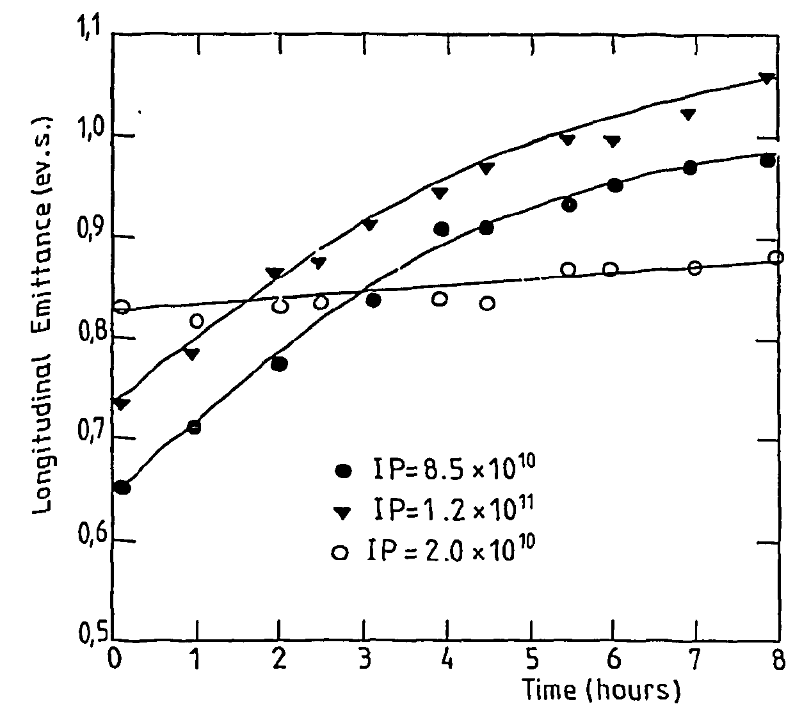}
\includegraphics[width=0.315\textwidth]{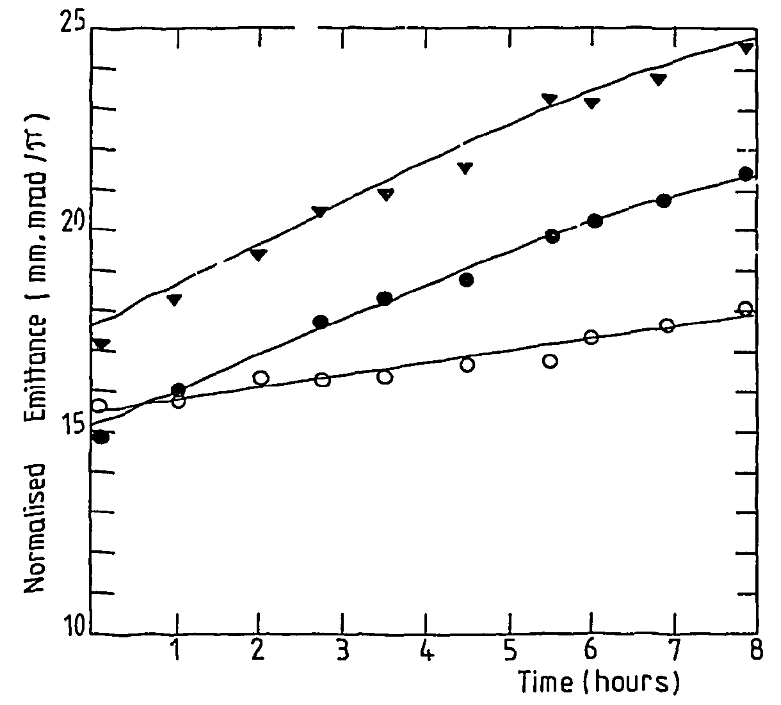}
\end{center}
\caption{Longitudinal and transverse emittance growth in the SPS \cite{gareyte1984}.
The left-hand plot shows longitudinal charge density measurements of proton and
antiproton bunches (left-hand and right-hand sides of the plot, respectively) made
every 30 minutes, with successive measurements aligned from the bottom to the top
of the plot.  There is a~larger increase in the bunch length for protons than antiprotons
because of the larger bunch population (of order
$10^{11}$ protons compared to $10^{10}$ antiprotons).  The middle and right-hand
plots show (respectively) the growth in longitudinal and transverse emittance over time
for proton bunches with different bunch populations.
\label{fig:gareyte1984}}
\end{figure}

Some of the earliest detailed observations of intrabeam scattering effects were made in the
CERN SPS in the early 1980's.  An example of the measurements is shown in
Fig.\,\ref{fig:gareyte1984} \cite{gareyte1984} (see also Fig.\,2 in
Ref.~\cite{piwinskibjorkenmtingwa2018}).  The left-hand plot shows measurements of the longitudinal
profile of a bunch of protons (left) and antiprotons (right) made at intervals of 30 minutes.
An increase in bunch length can be seen in both cases, but is more evident for the proton
bunch because of the relatively high population of the proton bunch (of order $10^{11}$
particles, compared to order $10^{10}$ particles for the antiproton bunch).  The central
plot shows the longitudinal emittance measured at intervals over a period of 8 hours, for
three proton bunches with different bunch populations.  The experimental measurements
(data points) are compared with theoretical predictions (lines).  The right-hand plot shows
corresponding measurements of the horizontal emittance. 

The measurements show good agreement between measurements and theoretical
predictions.  It is worth noting that other possible sources of emittance growth were
excluded: growth rates from gas scattering, for example, were estimated to be an
order of magnitude lower than those observed in the~machine under the given
conditions.  It can also be seen in Fig.\,\ref{fig:gareyte1984} that the growth rates
fall off as the~bunch emittance increases: this is expected for IBS, where the scattering
rates are reduced for lower particle density, but would not be expected for gas scattering.

\begin{figure}[t]
\begin{center}
\includegraphics[width=0.9\textwidth]{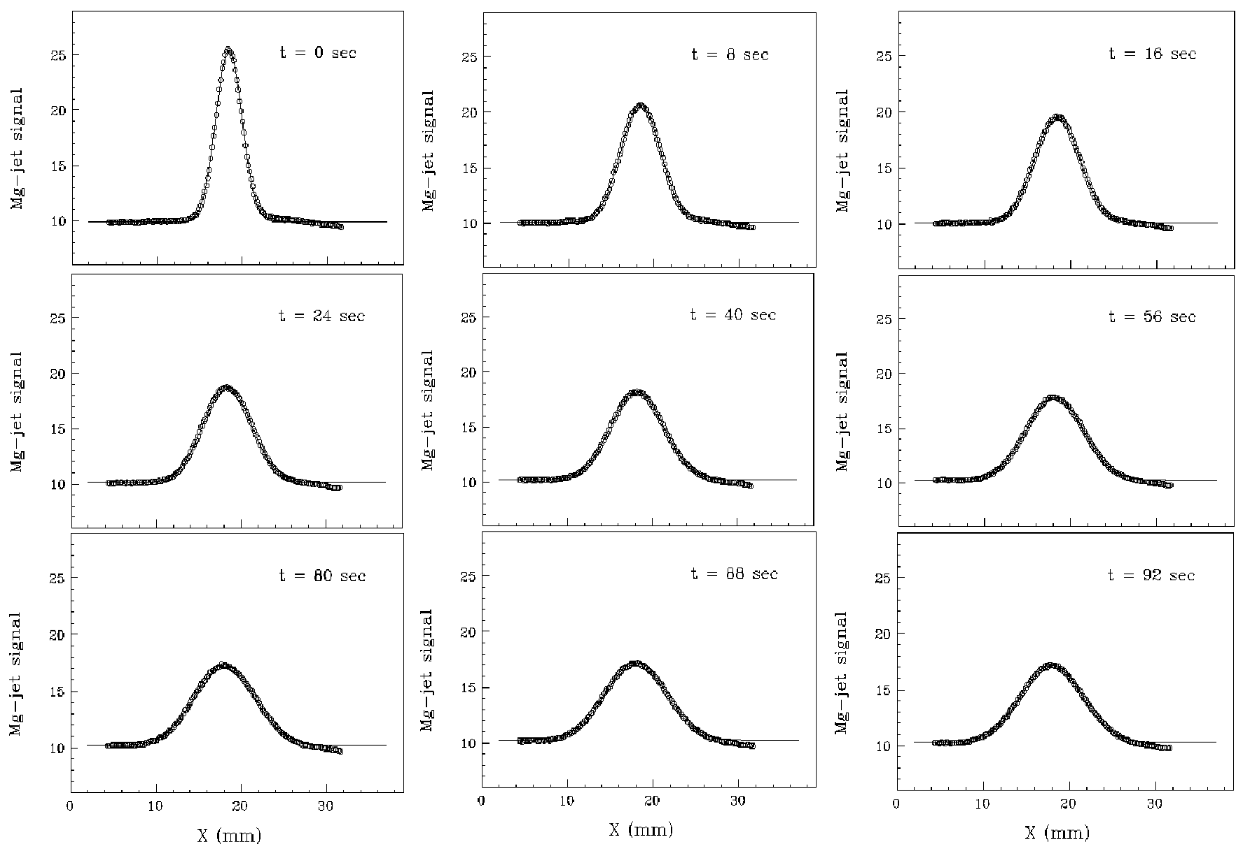}
\end{center}
\caption{Transverse emittance growth in CELSIUS \cite{raohermansson2000}.
Each plot shows the charge density in a 400\,MeV proton bunch in the storage ring
as a function of horizontal position within the bunch.  The protons are first accumulated
and then accelerated to 400\,MeV with beam cooling turned on.  For purposes of the
measurements shown, the~cooling is turned off after the bunch has reached equilibrium
at 400\,MeV, following which measurements are made at intervals for a total time of 92 seconds.
Circles show measured data, and the lines show Gaussian fits.
\label{fig:raohermansson2000}}
\end{figure}

Another example of emittance growth from intrabeam scattering is shown in
Fig.\,\ref{fig:raohermansson2000}.  In this case, measurements of the horizontal beam
size in a 400\,MeV proton beam in CELSIUS are presented.  The~bunch population is similar
to that shown in Fig.\,\ref{fig:gareyte1984} for the SPS, but in CELSIUS there is a clear
increase in the emittance on a timescale of seconds, rather than tens of minutes for the
SPS.  The reason for the~much larger emittance growth rate in CELSIUS is the beam energy:
whereas the SPS measurements were performed at 270\,GeV, the CELSIUS measurements
were made at 400\,MeV.  The growth rates also depend on the beam size and lattice functions,
but the significant difference between the two cases here is the difference in beam energy.

\begin{figure}[t]
\begin{center}
\includegraphics[width=0.32\textwidth]{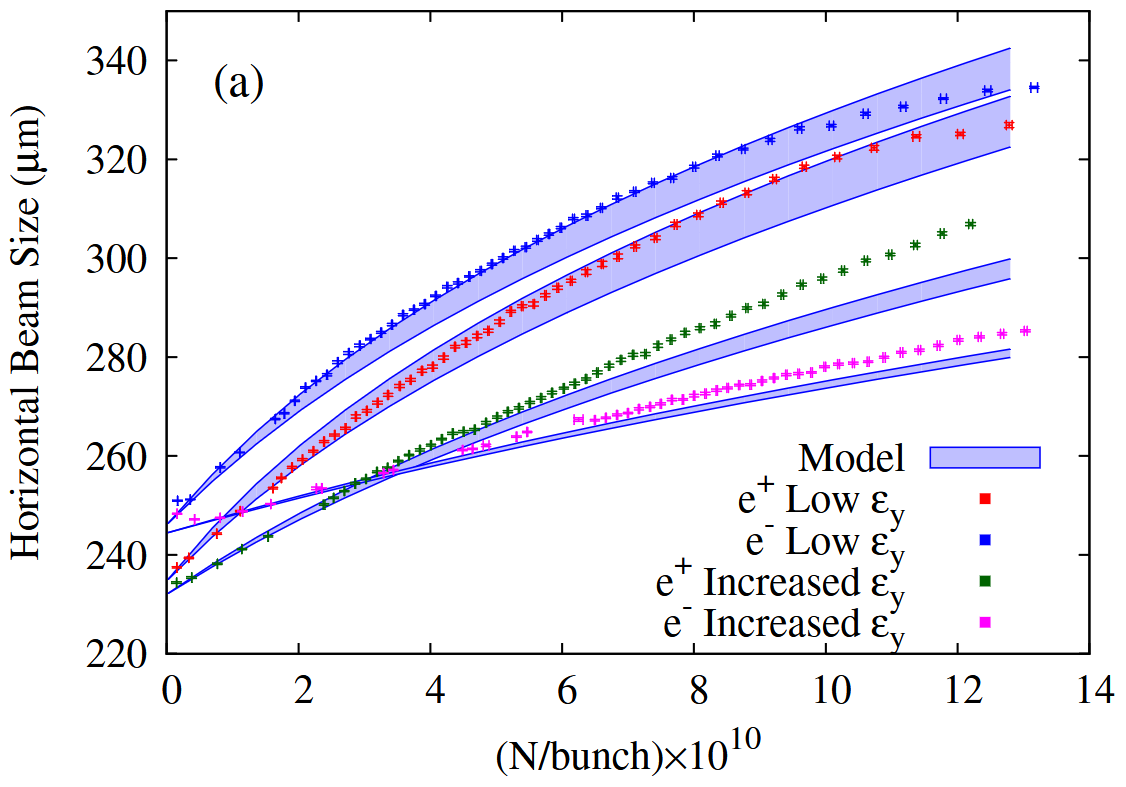}
\includegraphics[width=0.32\textwidth]{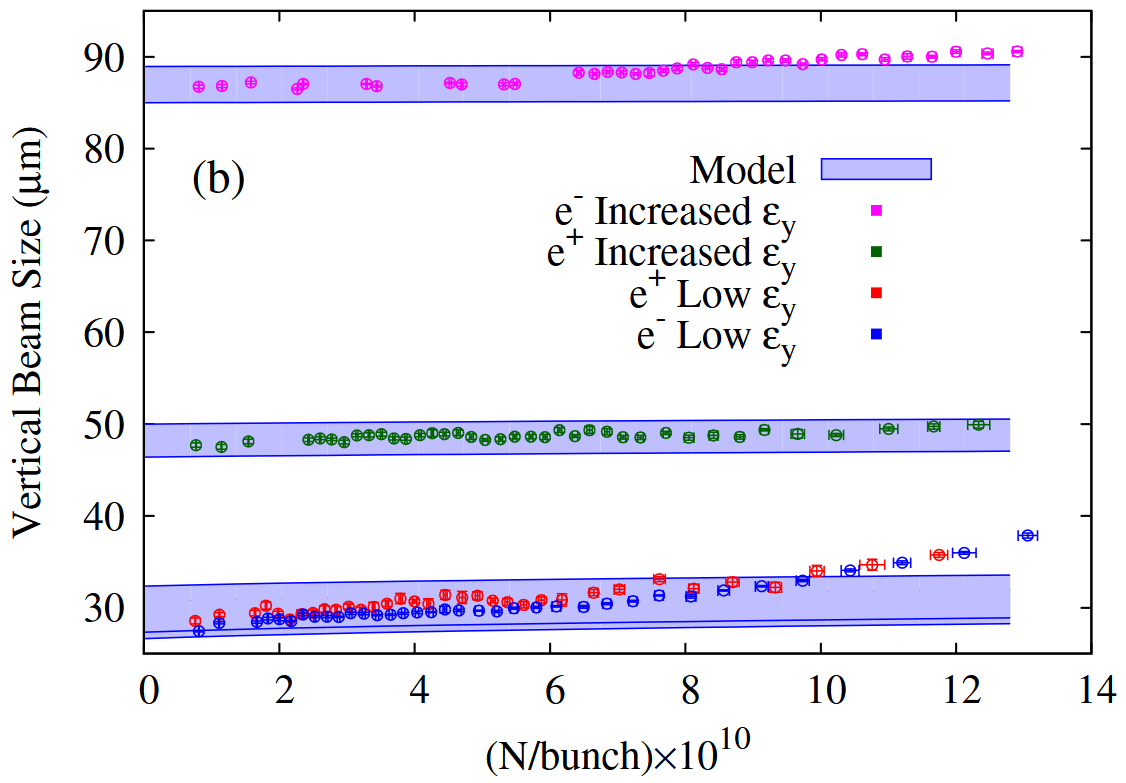}
\includegraphics[width=0.32\textwidth]{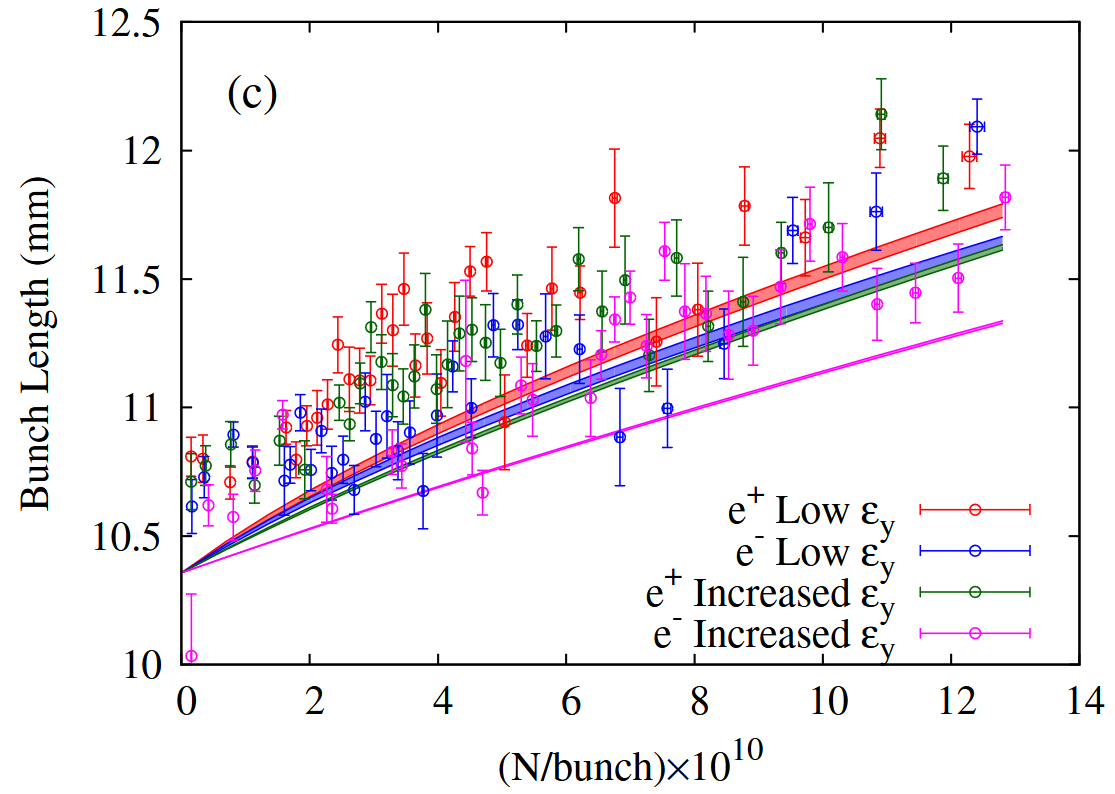}
\end{center}
\caption{Longitudinal and transverse emittance growth in the CESR storage ring
\cite{ehrlichman2013}.
Plots show the equilibrium horizontal (left) and vertical (middle) beam sizes and bunch
length (right) as functions of bunch population for electron and positron bunches.
Points show measurements and bands show theoretical predictions, taking into
account uncertainties in the beam conditions.  The vertical emittance can be controlled
by adjusting the coupling using skew quadrupoles.  Lower vertical emittances lead
to larger equilibrium horizontal emittance and bunch length because of the effects of intrabeam
scattering.
\label{fig:ehrlichman2013}}
\end{figure}

As a final example, measurements of IBS emittance growth are shown (for a very different
regime) in Fig.\,\ref{fig:ehrlichman2013} for the Cornell electron storage ring, CESR.
As already mentioned, in electron storage rings the~emittance growth rates from IBS are
generally much slower than the damping rates from synchrotron radiation.  However, the
very low vertical emittance that was achieved in CESR made it possible to observe an
increase in the equilibrium emittances as a function of bunch charge: the equilibrium is
determined by a balance between radiation damping, quantum excitation (again from
synchrotron radiation) and intrabeam scattering.  If the charge density in a bunch is
high enough, then the IBS emittance growth rates may be fast enough to have an
observable impact on the equilibrium emittance in an electron storage ring.  

\section{Theoretical models and growth-rate formulae}

\subsection{A simple gas model\label{sec:simplegasmodel}}

As a first step towards a theoretical analysis of intrabeam scattering, we can
consider a model in which the particles within a bunch in an accelerator are
represented as atoms or molecules in an ideal gas.  This is clearly a greatly
simplified picture of the system: it is not sufficiently accurate to provide a basis for
quantitative estimates of IBS growth rates, but it can give some insight into the
very different effects of IBS in storage rings operating below and above transition.

\begin{figure}[t]
\begin{center}
\includegraphics[width=0.95\textwidth]{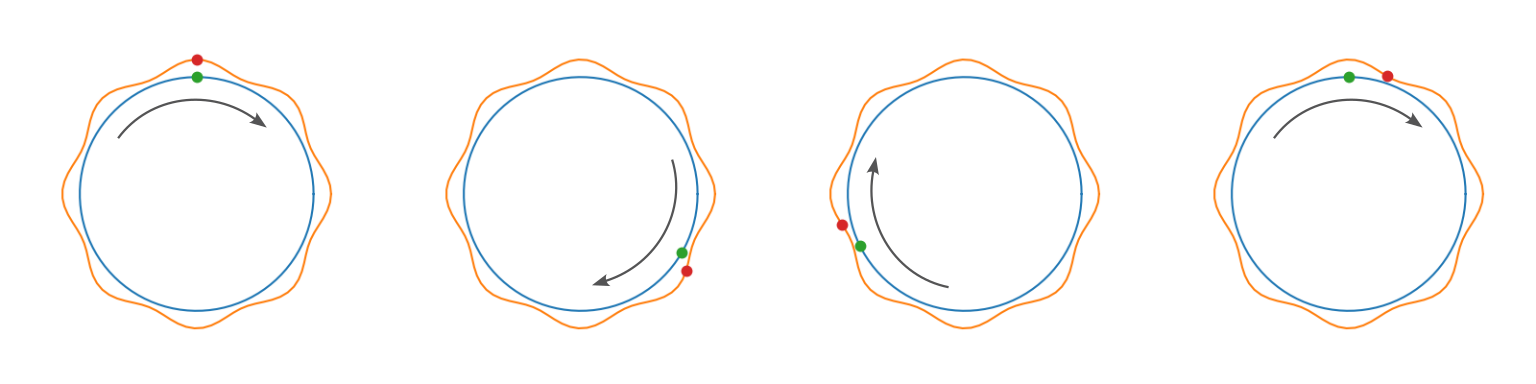} \\
(a) Below transition. \\
\includegraphics[width=0.95\textwidth]{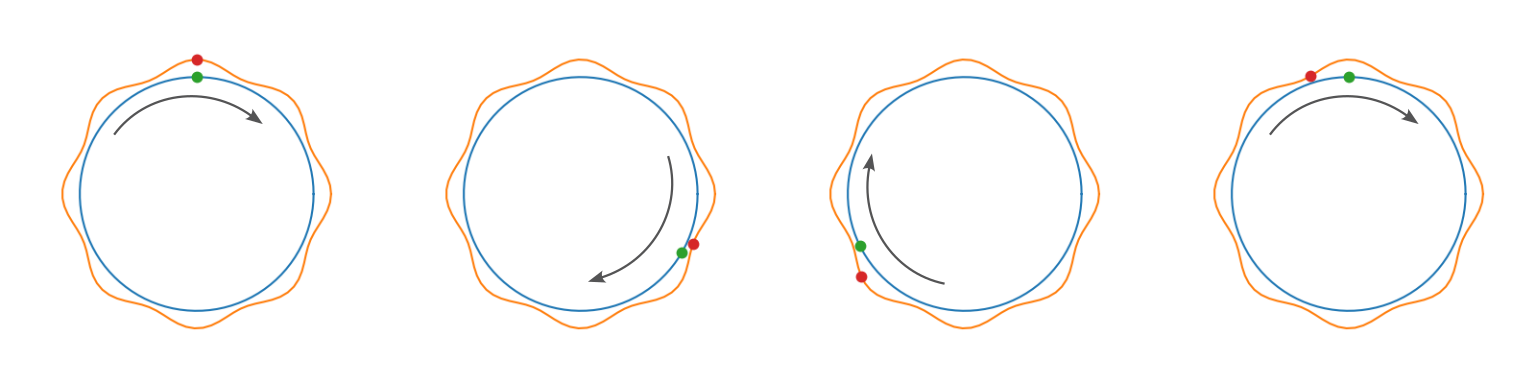} \\
(b) Above transition. \\
\end{center}
\caption{Longitudinal motion of particles in a storage ring (a) below transition and (b) above
transition.  In both cases, a higher-energy particle (represented by the red dot on the orange
outer path) follows a longer trajectory than a lower-energy particle (green dot on the blue
inner path). Below transition, the higher velocity associated with higher energy more than
compensates the increase in path length over one turn, so the revolution period falls with
increasing energy.  Above transition, particles are relativistic so that the increase
in velocity with energy is negligible, and because of the larger path length, higher-energy
particles take longer to complete each turn than lower-energy particles.
\label{fig:transition}}
\end{figure}

Consider first the case of a hadron storage ring operating below transition.  In most
storage rings, higher energy particles follow a dispersive trajectory that generally has a
larger circumference than the~nominal trajectory for particles with the design energy.
However, below transition the beam energy is low enough that the particle velocities
are not ultra-relativistic: a higher energy particle then has a higher velocity that more
than compensates the increase in the length of its trajectory over one turn around the~storage ring.  Hence, the revolution period of a particle \emph{decreases} as the energy
of the particle \emph{increases}: this is illustrated in Fig.\,\ref{fig:transition}\,(a).

In a storage ring below transition, particles scatter off each other in a way analogous to 
collisions between particles in an ideal gas.  One important difference is that while
the forces between particles in an ideal gas can be considered to be short-range
(as long as the particles carry no net electric charge) the~Coulomb force between
charged particles in an accelerator acts over relatively long distances.  A~further
difference is that an ideal gas is normally held in a container with rigid walls, and
particles move freely within the container until they collide with a wall or with another
particle.  In the case of particles in an~accelerator, we can represent the particles
as moving within a potential well provided by the~transverse and longitudinal
focusing from the magnets and RF cavities in the storage ring lattice.

Putting these differences aside, the important process for intrabeam scattering,
namely the transfer of momentum between the degrees of freedom through
elastic collisions between pairs of particles, is essential similar in both cases.
Let us further simplify the system by considering just longitudinal motion and
transverse motion in one dimension (either horizontal or vertical).  Assuming
that the potential well from the longitudinal and transverse focusing forces is
quadratic, the energy of a particle with mass $m$ may be written (in the rest
frame of the bunch, in a storage ring below transition):
\begin{equation}
E^\mathrm{(below)}_\mathrm{particle} = \frac{1}{2} m v_x^2 + \frac{1}{2} k_x x^2 + \frac{1}{2} m v_z^2 + \frac{1}{2} k_z z^2,
\label{particleenergybelowtransition}
\end{equation}
where $x$ and $z$ are the horizontal and longitudinal coordinates (respectively)
with corresponding velocities $v_x$ and $v_z$, and the transverse and longitudinal
focusing strengths are characterised by the constants $k_x$ and $k_z$.  Given this
expression for the energy of a particle and the fact that particles exchange energy
and momentum through elastic collisions, we can apply conservation of energy, so that:
\begin{equation}
m \langle v_x^2 \rangle + k_x \langle x^2 \rangle + m \langle v_z^2 \rangle + k_z \langle z^2 \rangle
= \frac{2E_\mathrm{total}}{N} = \textrm{constant},
\label{conservationofenergy}
\end{equation}
where the brackets $\langle \, \rangle$ indicate an average over all particles
in the bunch, there are $N$ particles in the~bunch and the total particle energy
(in  the rest frame of the bunch) is $E_\mathrm{total}$.

With some fairly general assumptions, we can derive relationships between the mean
square coordinates $\langle x^2 \rangle$, $\langle z^2 \rangle$ and the mean square
velocities $\langle v_x^2 \rangle$, $\langle v_z^2 \rangle$.  For a single oscillator in one
dimension, we can write the equation of motion:
\begin{equation}
x = A\sin(\omega t + \psi),
\end{equation}
where $\omega = \sqrt{k_x/m}$, and the amplitude $A$ and initial phase $\psi$ are constants.
It follows that the averages of $x^2$ and $v_x^2$ over a long time interval (i.e.~long compared to
the oscillation period) are:
\begin{equation}
\langle x^2 \rangle_t = \frac{1}{2}A^2, \qquad
\langle v_x^2 \rangle_t = \frac{1}{2}\omega^2 A^2,
\end{equation}
and hence:
\begin{equation}
k_x \langle x^2 \rangle_t = m \langle v_x^2 \rangle_t.
\label{timeaveragesho}
\end{equation}
Given a large collection of particles with random amplitudes and initial phase angles
(so that each particle at any moment is at a different phase in its cycle of oscillation)
we expect the same relationship to hold between the mean square coordinate and
mean square velocity with the mean values now taken over all particles at a given
time\footnote{The fact that we can substitute averages over all particles for averages over
time in equation (\ref{timeaveragesho}) follows from the ergodic theorem.}.
Extending the argument to include (separately) the longitudinal motion, we then have:
\begin{equation}
m \langle v_x^2 \rangle = k_x \langle x^2 \rangle, \quad
\textrm{ and } \quad
m \langle v_z^2 \rangle = k_z \langle z^2 \rangle.
\label{equipartitionofenergy}
\end{equation}
In the context of accelerator beam dynamics, Eqs.~(\ref{equipartitionofenergy})
are the conditions for a \emph{matched} distribution in a~storage ring: the shape of the
distribution of a bunch of particles in a storage ring will not change over time if the
distribution is correctly matched to the focusing in the lattice. In terms of statistical
mechanics, each of the equations in Eq.~(\ref{equipartitionofenergy}) is essentially a
statement of equipartition of energy applied to a collection of particles in equilibrium
for the transverse and longitudinal motion separately.

Assuming that the particle velocities are not correlated with the coordinates\footnote{Since
$\langle x p_x \rangle = -\alpha_x \varepsilon_x$, where $p_x$ is the particle momentum
(in the laboratory frame), $\alpha_x = -\frac{1}{2} d\beta_x/ds$ is the Twiss alpha function
(related to the derivative of the beta function $\beta_x$), assuming that
$\langle xv_x \rangle = 0$ implies that $\alpha_x = 0$ and hence $\beta_x = $constant.} so
that $\langle xv_x \rangle = \langle zv_z \rangle = 0$, the (geometric) emittances can be written:
\begin{equation}
\varepsilon_x = \frac{1}{P_0}\sqrt{\langle x^2 \rangle \langle p_x^2 \rangle}, \qquad
\varepsilon_z = \frac{1}{P_0}\sqrt{\langle z^2 \rangle \langle p_z^2 \rangle},
\label{restframeemittances}
\end{equation}
where $p_x = mv_x$, $p_z = mv_z$, and $P_0$ is the reference momentum (the
momentum of a particle with the~design energy of the beam in the laboratory
frame).  Combining Eqs.~(\ref{conservationofenergy}),
(\ref{equipartitionofenergy}) and (\ref{restframeemittances}) we find:
\begin{equation}
P_0 \sqrt{\frac{k_x}{m}} \varepsilon_x + P_0 \sqrt{\frac{k_z}{m}} \varepsilon_z = \frac{E_\mathrm{total}}{N} = \textrm{constant}.
\label{conservationofenergy2}
\end{equation}
Equation (\ref{conservationofenergy2}) must be satisfied even in the presence of
any effect, such as intrabeam scattering, that can cause an exchange of energy between
the transverse and longitudinal motion.  An increase in energy in the transverse
direction will increase the transverse emittance $\varepsilon_x$: but then we see that
the longitudinal emittance $\varepsilon_z$ must be reduced.  Starting from a distribution
with arbitrary transverse and longitudinal emittances, the bunch will move towards an
equilibrium in which equipartition of energy is satisfied even taking into account the
exchange in energy and momentum between the transverse and longitudinal motion:
\begin{equation}
m \langle v_x^2 \rangle = k_x \langle x^2 \rangle = m \langle v_z^2 \rangle = k_z \langle z^2 \rangle.
\label{equipartitionofenergy2}
\end{equation}
The above arguments are readily extended to three degrees of freedom.  In a storage ring
below transition, conservation of energy means that although IBS causes a transfer of energy
and emittance between the~degrees of freedom, the emittances will always satisfy:
\begin{equation}
\sqrt{k_x} \varepsilon_x + \sqrt{k_y} \varepsilon_y + \sqrt{k_z} \varepsilon_z = \textrm{constant}.
\label{conservationofenergy3}
\end{equation}
Below transition, the beam will reach an equilibrium in which the emittances satisfy
the equipartition of energy:
\begin{equation}
\sqrt{k_x} \varepsilon_x = \sqrt{k_y} \varepsilon_y = \sqrt{k_z} \varepsilon_z.
\label{equipartitionofenergy3}
\end{equation}

In a storage ring above transition, the situation is very different.  In this regime,
particles now have ultra-relativistic velocities in the laboratory frame.  A particle
with energy above the reference energy will again follow a dispersive trajectory with
a circumference that is longer than the nominal trajectory; but because the
particle is ultra-relativistic there is negligible difference in the velocity compared
to the~velocity of a particle with the reference energy.  As a result, the orbital
period for a particle will \emph{increase} with an increase in energy: see Fig.~\ref{fig:transition}\,(b).
In the rest frame of the bunch, this is as if an increase in the~longitudinal velocity
was associated with a \emph{reduction} in energy.  In the transverse directions there
is no change compared to the case of a storage ring below transition: an increase in
velocity in a transverse direction will still be associated with an increase in energy.

Despite the change in the nature of the longitudinal motion, we can still apply the
ideal gas model to understand the properties of the bunch, as long as we treat particles
as having \emph{negative mass} for their longitudinal motion: a force acting in the $+z$
direction will lead to a change in velocity towards the $-z$ direction.  This implies that
to maintain a bunched beam, the focusing in the longitudinal direction must also now
be negative ($k_z < 0$).  This is of course the situation in a real storage ring: below
transition, the~RF cavities must be phased so that bunches cross the cavities on a rising
slope of the voltage (particles with higher energy with lower revolution period arrive
earlier at a cavity and see a lower RF voltage); but above transition, the cavities must be
phased so that bunches cross the cavities on a falling slope of the~RF voltage.  In terms
of the potential energy of particles (viewed in the rest frame of the bunch), the~larger
the longitudinal co-ordinate $z$, the more negative the potential energy of the particle.

Taking these differences into account, for a storage ring above transition the expression
corresponding to Eq.~(\ref{particleenergybelowtransition}) for the energy of a
particle is:
\begin{equation}
E^\mathrm{(above)}_\mathrm{particle} = \frac{1}{2} m v_x^2 + \frac{1}{2} k_x x^2 - \frac{1}{2} |m| v_z^2 - \frac{1}{2} |k_z| z^2.
\end{equation}
Equation (\ref{equipartitionofenergy}) for the shape of the distribution treating
each degree of freedom independently is still valid; and we can still use the definitions
Eq.~(\ref{restframeemittances}) for the emittances.  But a larger longitudinal emittance
corresponds to a more negative (i.e.~lower) energy, and then following the same reasoning
as for the case of a storage ring below transition, conservation of energy now leads to:
\begin{equation}
\sqrt{k_x} \varepsilon_x + \sqrt{k_y} \varepsilon_y - \sqrt{|k_z|} \varepsilon_z = \textrm{constant}.
\label{conservationofenergy4}
\end{equation}
The minus sign before the final term on the left hand side of this equation has a
significant consequence: \emph{below} transition conservation of energy required
at least one emittance to shrink in order for another emittance to grow, but \emph{above}
transition all three emittances can grow indefinitely.  In a storage ring operating
above transition, there is no beam equilibrium (i.e.~there is no condition that would
correspond to Eq.~(\ref{equipartitionofenergy3}) in the case of a storage ring
below transition).

Of course, intrabeam scattering simply provides a means for exchange of
momentum of particles between the different degrees of freedom, and the simple
gas model we have used in the above arguments cannot provide a fully detailed
description of the effects of IBS.  The model does offer some insight into the
very different impact of IBS in a storage ring above transition compared to the case
of a storage ring below transition; but to make any precise predictions of IBS effects
in either case, the details of the~scattering process must be considered.

\subsection{Piwinski formulae for IBS growth rates}

The fundamental process that takes place when two particles within a single bunch
in a hadron accelerator collide with each other is Rutherford scattering.  Formulae for
the rates of change of the bunch emittances as a result of the scattering process can
be derived starting from a formula for the rate of scattering of particles from one
region of momentum space to another.  This depends on the phase space density
of particles (the number of particles per unit volume of co-ordinate space and
momentum space), the~energies of the particles, and the scattering amplitude.
The scattering amplitude describes the interaction between particles at the
quantum level.

Consider a collection of particles in a small volume of co-ordinate space $d^3\!x$.
Particles within this volume will have a range of momenta.  In the laboratory frame,
the rate of scattering of
pairs of particles from regions of momentum space $d^3p_1$, $d^3p_2$ (for the two
members of the pair) to regions $d^3p_1^\prime$, $d^3p_2^\prime$ can be written:
\begin{eqnarray}
\frac{d \mathcal{P}}{dt} & = & \frac{1}{2} \rho(x, p_1) \rho(x, p_2)
\, d^3x \, \frac{d^3p_1}{\gamma_1} \frac{d^3p_2}{\gamma_2} \cdot
|\mathcal{M}|^2 \cdot
\frac{d^3p_1^\prime}{\gamma_1^\prime} \frac{d^3p_2^\prime}{\gamma_2^\prime} \nonumber \\
& & \quad \times 
\frac{\delta^{(3)}(p_1^\prime + p_2^\prime - p_1 - p_2) \, \delta(E_1^\prime + E_2^\prime - E_1 - E_2)}{(2\pi)^2},
\label{scatteringrate}
\end{eqnarray}
where $\rho(x,p)$ is the phase space density of the beam, and
$\mathcal{M}$ is the scattering amplitude.  Note that this formula for the~scattering rate is given in the laboratory frame where the two particles involved
in the~scattering process have energies $E_1$ and $E_2$.  For protons, the~scattering amplitude $\mathcal{M}$ is given by:
\begin{equation}
\mathcal{M} = \frac{4\pi \alpha}{q^2},
\label{scatteringamplitude}
\end{equation}
where $\alpha \approx 1/137$ is the fine structure constant, and $q$ is the
change in four-momentum of either particle during the collision.  The rate of
change of the mean value $\langle g \rangle_\mathrm{PS}$ of any function $g(x,p)$ of the
phase space variables is given by:
\begin{equation}
\frac{d}{dt}\langle g \rangle_\mathrm{PS} = \int_\mathrm{PS} \frac{d\mathcal{P}}{dt} \Big(
g(x, p_1^\prime) - g(x, p_1) + g(x, p_2^\prime) - g(x, p_2) \Big),
\label{phasespacerateofchange}
\end{equation}
where the brackets $\langle \cdot \rangle_\mathrm{PS}$ indicate an average over
phase space, and the integral is taken over all phase space.

Equations (\ref{scatteringrate}), (\ref{scatteringamplitude}), and (\ref{phasespacerateofchange})
provide the basic formulae needed to calculate the rate of change of emittance
of a bunch in an accelerator as a result of IBS.  The calculation is not
straightforward, however, because of the nature of the integrals involved even
for ``simple'' Gaussian beams.  Furthermore, the~phase space distribution of a bunch
will change according to the variation of the lattice functions as the~bunch moves
along an accelerator beam line.  This means that in a storage ring, further integrals
must be performed to average the growth rates around the circumference of the ring.
Depending on the~details of how the calculation is performed, including any assumptions
or approximations (e.g.~assuming constant values for the beta functions) different
expressions can be obtained for the growth rate formulae.

The earliest complete analysis of the IBS growth rates was performed by Piwinski
\cite{piwinski1974}.  Piwinski's original calculation assumed a Gaussian beam distribution
and neglected beta function gradients (which lead to correlations between co-ordinates
and momenta in phase space).  The calculations yield formulae for the IBS growth rates
$1/\tau_i$, defined by\footnote{The factor of 2 on the right hand side of Eq.~(\ref{growthratedefinition}) is included by analogy with the conventional definition
of the~synchrotron radiation damping times in electron storage rings: since the
beam size is proportional to the square root of the emittance, if the~emittance damps
with exponential time constant $2\tau$, then the beam size damps with time constant $\tau$.}:
\begin{equation}
\frac{d\varepsilon_i}{dt} = \frac{2\varepsilon_i}{\tau_i},
\label{growthratedefinition}
\end{equation}
where $i = x,y,z$ for the horizontal, vertical and longitudinal emittance and growth rate,
respectively.  In Piwinski's theory, the formulae for the growth rates can be written:
\begin{eqnarray}
\frac{1}{\tau_x} & = & A \left\langle f \! \left( \frac{1}{\mytilde{a}}, \frac{\mytilde{b}}{\mytilde{a}}, \frac{\mytilde{q}}{\mytilde{a}} \right) +
\frac{\eta_x^2 \sigma_h^2}{\beta_x \varepsilon_x} f(\mytilde{a},\mytilde{b},\mytilde{q}) \right\rangle_{\!\!\mathrm{C}},
\label{ibspiwinskix} \\
\frac{1}{\tau_y} & = & A \left\langle f \! \left( \frac{1}{\mytilde{b}}, \frac{\mytilde{a}}{\mytilde{b}}, \frac{\mytilde{q}}{\mytilde{b}} \right) +
\frac{\eta_y^2 \sigma_h^2}{\beta_y \varepsilon_y} f(\mytilde{a},\mytilde{b},\mytilde{q}) \right\rangle_{\!\!\mathrm{C}},
\label{ibspiwinskiy} \\
\frac{1}{\tau_z} & = & A \left\langle \frac{\sigma_h^2}{\sigma_\delta^2} f(\mytilde{a},\mytilde{b},\mytilde{q}) \right\rangle_{\!\!\mathrm{C}}. \label{ibspiwinskiz}
\end{eqnarray}
The brackets $\langle \cdot \rangle_{\mathrm{C}}$ denote an average around the
circumference of the ring.  The factor $A$ describes the~scaling of the growth
rates with energy and the phase space density, and is given by:
\begin{equation}
A = \frac{\pi r_0^2 c N_b}{8\gamma_0 \Gamma},
\label{ibsafactor}
\end{equation}
where $r_0 = e^2 / (4\pi \permittivity m c^2)$ is the classical
radius of the particle (with charge $e$ and mass $m$),
$N_b$ is the~bunch population,
$\gamma_0$ is the relativistic factor for the beam, and
$\Gamma = (2\pi)^3 \varepsilon_{nx} \varepsilon_{ny} \varepsilon_{nz}$ is the phase
space volume of a bunch with \emph(normalised) emittances $\varepsilon_{nx}$,
$\varepsilon_{ny}$, $\varepsilon_{nz}$.

\begin{figure}
\begin{center}
\includegraphics[width=0.5\textwidth]{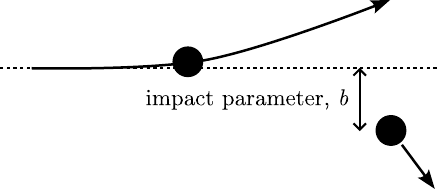}
\caption{In a collision between two particles, observed in a frame in which
one particle is initially at rest, the impact parameter ($b$ in the figure) measures
the perpendicular distance between the initial trajectory of the moving particle
and the initial position of the stationary particle.
\label{figimpactparameter}}
\end{center}
\end{figure}

\begin{figure}[t]
\begin{center}
\includegraphics[width=0.98\textwidth]{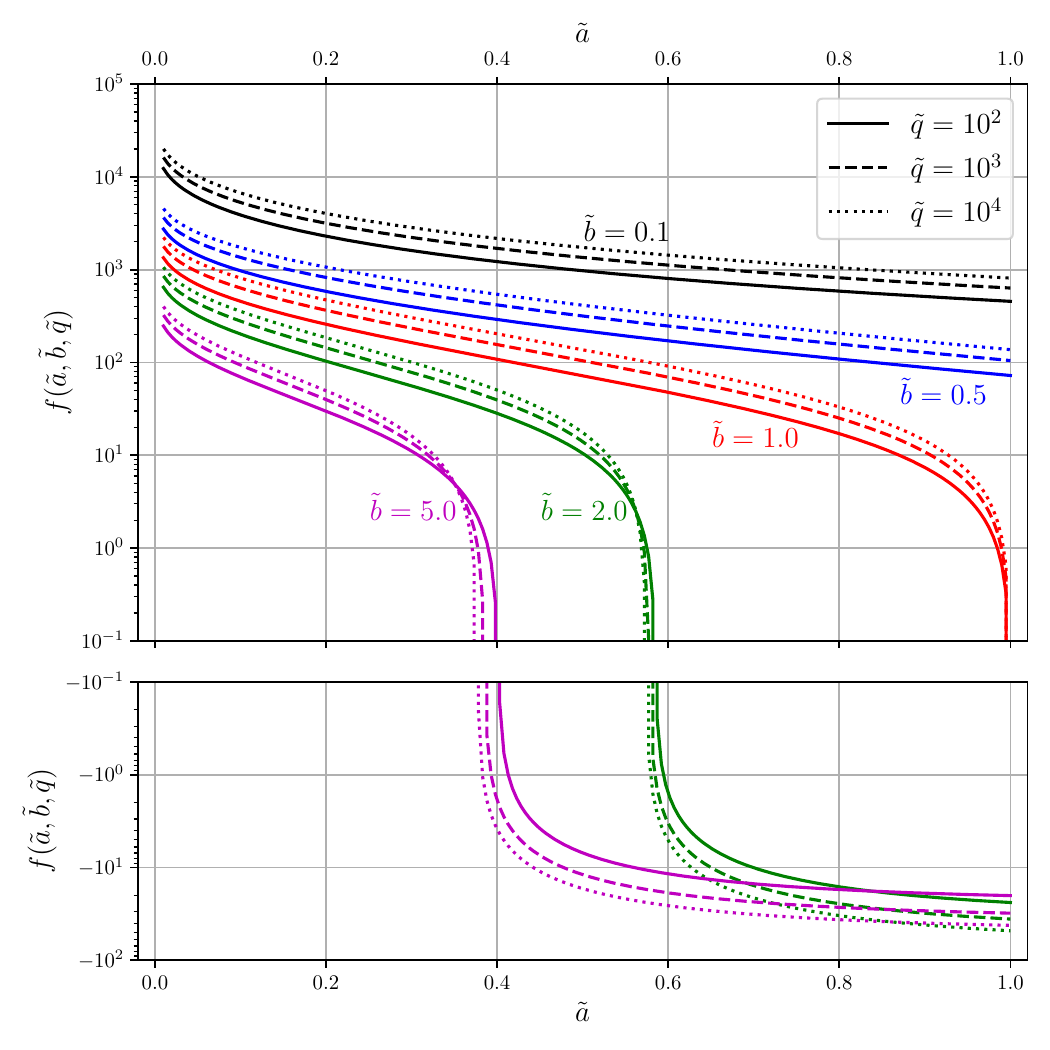}
\caption{Piwinski function $f$, defined by Eq.~(\ref{ibspiwinskiffunction}).
The Piwinski function characterises the dependence of the~growth rates on the
beam properties, as expressed in Eqs.~(\ref{ibspiwinskix}), (\ref{ibspiwinskiy})
and (\ref{ibspiwinskiz}).
The parameters $\tilde{a}$ and $\tilde{b}$, defined in Eq.~(\ref{ibspiwinskiq}), 
characterise (broadly) the dependence of the IBS scattering rate on the horizontal
and vertical beta functions (respectively).  The parameter $\tilde{q}$ includes the
dependence on the ratio of the maximum to minimum impact parameter in any
collision between particles.  Note the strong dependence of the Piwinski function
on $\tilde{a}$ and $\tilde{b}$, and the relatively weak dependence on $\tilde{q}$.
\label{figpiwinskif}}
\end{center}
\end{figure}

The parameters $\mytilde{a}$, $\mytilde{b}$, $\mytilde{q}$ are
defined: 
\begin{equation}
\mytilde{a} = \frac{\sigma_h}{\gamma_0} \sqrt{\frac{\beta_x}{\varepsilon_x}},
\quad 
\mytilde{b} = \frac{\sigma_h}{\gamma_0} \sqrt{\frac{\beta_y}{\varepsilon_y}},
\quad 
\mytilde{q} = \beta_0 \sigma_h \sqrt{\frac{2b_\mathrm{max}}{r_0}},
\label{ibspiwinskiq}
\end{equation}
where:
\begin{equation}
\frac{1}{\sigma_h^2} = \frac{1}{\sigma_\delta^2}
+ \frac{\eta_x^2}{\beta_x \varepsilon_x}
+ \frac{\eta_y^2}{\beta_y \varepsilon_y},
\label{ibspiwinskisigmah}
\end{equation}
$\sigma_z$ and $\sigma_\delta$ are the bunch length and energy spread,
$\beta_x$, $\beta_y$ are the horizontal and vertical beta functions, $\eta_x$,
$\eta_y$ are the horizontal and vertical dispersion, and $b_\mathrm{max}$
is the maximum impact parameter in a scattering event, with the impact
parameter defined as shown in Fig.~\ref{figimpactparameter}. The ``Piwinski function''
$f(\mytilde{a},\mytilde{b},\mytilde{q})$, shown for selected values of the
parameters in Fig.~\ref{figpiwinskif}, is given by:
\begin{equation}
f(\mytilde{a},\mytilde{b},\mytilde{q}) = 8\pi \int_0^1
\left( 2 \ln \! \left( \frac{\mytilde{q}}{2}
\left( \frac{1}{\mytilde{P}} + \frac{1}{\mytilde{Q}} \right) \right) - 0.577\ldots \right)
 \frac{1 - 3u^2}{\mytilde{P}\mytilde{Q}} \, du,
\label{ibspiwinskiffunction}
\end{equation}
where:
\begin{equation}
\mytilde{P}^2 = \mytilde{a}^2 + (1 - \mytilde{a}^2) u^2, \qquad
\mytilde{Q}^2 = \mytilde{b}^2 + (1 - \mytilde{b}^2) u^2.
\end{equation}

The parameters $\tilde{a}$ and $\tilde{b}$ characterise the dependence of the
scattering rate on the horizontal and vertical beta functions (respectively); but
also include effects of coupling between transverse and longitudinal motion through
$\sigma_h$ and the dispersion functions.  The dependence of the transverse growth
rates on the dispersion arises from the fact that, because the closed orbit in a storage
ring depends on the energy of the particle, a sudden change in the energy of a particle
will lead to a change in betatron amplitude.  The effect is analogous to quantum
excitation in electron storage rings: a particle that emits a photon (in a dipole
or insertion device) at a location of non-zero dispersion will undergo a change
in betatron amplitude as a result of the change in closed orbit related to the
change in energy of the particle.

One of the difficulties in deriving formulae for the emittance growth rates
resulting from IBS is the~fact that the scattering amplitude in Eq.~(\ref{scatteringamplitude})
diverges when the change $q$ in the four-momentum of a particle involved in
a collision is such that $q^2 = 0$.  This occurs when there is no change in energy
of either particle in the collision, which can happen in two cases: first, where
the particles collide head-on (impact parameter equal to zero), and second, in the
limit of large impact parameter.  To avoid divergence of the formulae for the
growth rates, it is necessary to set upper and lower limits on the impact parameter.
Unfortunately, there is no completely rigorous way of specifying the limits given
the model on which the~theory is based.  For example, if the impact parameter is
very large, then this implies that at least one of the~particles is a~large
distance from the core of the bunch: in a Gaussian distribution, although the
density of particles falls with increasing distance from the centre, mathematically
it never falls completely to zero.  The reduction in particle density with
increasing impact parameter balances the divergence of the~scattering amplitude
to some extent, but is not sufficient on its own to prevent divergence of the~formulae for the~emittance growth rates.

To address the issues associated with the divergence of the scattering amplitude,
it is usual to take a pragmatic approach and set upper and lower limits on
the impact parameter that reflect the nature of the particles and the distribution
of particles in a bunch.  Conventionally, the lower limit is specified to be the
classical radius of the particle, $r_0 = e^2/(4\pi \permittivity m c^2)$, where
$e$ is the charge on the particle and $m$ is the~mass.  The upper limit is taken
to be the smallest dimension of the bunch: this is usually the~vertical beam size,
so that $b_\mathrm{max} = \sqrt{\beta_y \varepsilon_y}$ in Eq.~(\ref{ibspiwinskiq}). Although these choices may seem rather arbitrary, fortunately
the scattering rates are fairly insensitive to the precise values chosen: the limits
enter the formulae for the scattering rates as a ratio in the expression for the
parameter $\tilde{q}$ in Eq.~(\ref{ibspiwinskiq}), and we see from Fig.~\ref{figpiwinskif}
that the Piwinski function $f(\tilde{a},\tilde{b},\tilde{q})$ has a relatively weak
dependence on $\tilde{q}$ (compared to its dependence on $\tilde{a}$ and
$\tilde{b}$).

\subsection{Derivatives of the lattice functions: Martini's formulae}
Piwinski's formulae for the IBS growth rates are expressed in terms of the lattice
functions (beta function and dispersion).  In principle, it is possible to take into
account variations in the lattice functions around a storage ring by calculating the
growth rates as a function of position.  However, Piwinski's formulae neglect effects
associated with the derivatives of the lattice functions with respect to position,
i.e.~dependence of the scattering rates on $\beta^\prime_{x,y}=-2\alpha_{x,y}$
and $\eta^\prime_x$. These quantities describe (respectively) the correlation between
the transverse coordinates and momenta of particles, and the dependence of the~transverse momentum of a particle following a closed orbit on the energy of the
particle. Since the~rate of momentum transfer between particles in IBS depends on
the distributions of particles in phase space ($\rho(x,p)$ in Eq.~(\ref{scatteringrate})), the values of the derivatives of the lattice functions can affect
the emittance growth rates. Piwinski's theory was generalised by Martini
\cite{martini1984} to take the derivatives of the lattice functions into account.
The corrections are generally quite small; we do not give the formulae here, but
refer the reader to the literature, e.g.~\cite{martini1984, martini2015, martini2016}.

\subsection{Bjorken--Mtingwa formulae}
Working independently from Piwinski, and following an alternative approach to
the analysis, Bjorken and Mtingwa \cite{bjorkenmtingwa1983} derived
expressions for the IBS growth rates that take a different form from the formulae
given in the previous section.  In the Bjorken--Mtingwa formalism, the growth
rates are given by:
\begin{eqnarray}
\frac{1}{T_i} & = & 4\pi A (\log)_\mathrm{BM} \times \nonumber \\
& & \left\langle
\int_0^\infty d\lambda \sqrt{\frac{\lambda}{\det(L + \lambda I)}}
\left[
\Tr(L_i) \Tr \! \left( \frac{1}{L + \lambda I} \right)
- 3\Tr \left( \frac{L_i}{L + \lambda I} \right)
\right]
\right\rangle_\mathrm{C}, \nonumber  \label{ibsgrowthratebm}
\end{eqnarray}
where $A$ is given by Eq.~(\ref{ibsafactor}), $I$ is the identity matrix, the
matrices $L_i$ are given by:
\begin{eqnarray}
L_x & = & \frac{\gamma_0 \sqrt{\beta_x \mathcal{H}_x}}{\varepsilon_x} \left(
\begin{array}{ccc}
\frac{1}{\gamma_0} \sqrt{\frac{\beta_x}{\mathcal{H}_x}} &  \sin (\varphi_x) & 0 \\
\sin (\varphi_x) & \gamma_0 \sqrt{\frac{\mathcal{H}_x}{\beta_x}} & 0 \\
0 & 0 & 0
\end{array}
\right), \\
L_y & = & \frac{\gamma_0 \sqrt{\beta_y \mathcal{H}_y}}{\varepsilon_y} \left(
\begin{array}{ccc}
0 & 0 & 0 \\
0 & \gamma_0 \sqrt{\frac{\mathcal{H}_y}{\beta_y}} & \sin (\varphi_y) \\
0 & \sin (\varphi_y) & \frac{1}{\gamma_0} \sqrt{\frac{\beta_y}{\mathcal{H}_y}}
\end{array}
\right), \\
L_z & = & \frac{\gamma_0^2}{\sigma_\delta^2} \left(
\begin{array}{ccc}
0 & 0 & 0 \\
0 & 1 & 0 \\
0 & 0 & 0
\end{array}
\right),
\end{eqnarray}
and $L$ is defined as:
\begin{equation}
L = L_x + L_y + L_z.
\end{equation}
$\mathcal{H}_x$ and $\mathcal{H}_y$ are the horizontal and vertical dispersion
invariants\footnote{The functions $\mathcal{H}_x$ and $\mathcal{H}_y$ are referred
to as ``invariants'' because outside of dipole magnets they take constant values as
functions of position along a beamline.}, defined in terms of the Twiss parameters
and the dispersion functions.  The horizontal dispersion invariant is given by:
\begin{equation}
\mathcal{H}_x = \gamma_x \eta_x^2 + 2 \alpha_x \eta_x \eta_{px} + \beta_x \eta_{px}^2.
\label{ibscurlyhfunction1}
\end{equation}
A similar expression holds for the vertical dispersion invariant $\mathcal{H}_y$.
The function $\varphi_x$ is defined by:
\begin{equation}
\tan (\varphi_x) = - \beta_x \frac{\eta_{px}}{\eta_x} - \alpha_x,
\label{ibscurlyphifunction}
\end{equation}
with a similar definition for $\varphi_y$.  The quantities $\mathcal{H}_x$ and $\phi_x$,
with their counterparts in the vertical direction, describe the amplitude and phase
of the dispersion at different points around a storage ring.  In particular, the horizontal
dispersion at any point in the ring can be written:
\begin{eqnarray}
\eta_x & = & \sqrt{\beta_x \mathcal{H}_x} \cos (\varphi_x), \\
\eta_{px} & = & - \sqrt{\frac{\mathcal{H}_x}{\beta_x}} \left( \sin (\varphi_x) + \alpha_x \cos (\varphi_x) \right). 
\end{eqnarray}
In this respect, the functions $\mathcal{H}_x$ and $\phi_x$ are analogous to the
betatron action and phase that describe betatron oscillations of particles moving around
the ring.

The quantity $(\log)_\mathrm{BM}$, sometimes known as the \emph{Coulomb log} is defined as:
\begin{equation}
(\log)_\mathrm{BM} = \ln \! \left( \frac{b_\mathrm{max}}{b_\mathrm{min}} \right),
\end{equation}
where $b_\mathrm{max}$ is the maximum impact parameter that can occur in the collision
between two particles in the beam, and $b_\mathrm{min}$ is the minimum impact parameter that
can occur. The values for these parameters are sometimes chosen as follows:
\begin{eqnarray}
b_\mathrm{max} & = & \mathrm{min}(\sigma_x, \sigma_y), \\
b_\mathrm{min} & = & r_0,
\end{eqnarray}
where $\mathrm{min}(\sigma_x, \sigma_y)$ is the smaller of the horizontal and vertical
beam size, and $r_0$ is the classical radius of the particles.

The expressions for the IBS growth rates in the Bjorken--Mtingwa formula
(\ref{ibsgrowthratebm}) take a form that looks very different from the expressions
for the growth rates in the Piwinski formulae (\ref{ibspiwinskix}), (\ref{ibspiwinskiy})
and (\ref{ibspiwinskiz}).  However, Bane \cite{bane2002} has shown that, with
certain assumptions, the Piwinski formulae and the~Bjorken--Mtingwa formulae are in
good agreement with each other.  This is discussed further in the following section,
where we consider approximations to the formulae for the IBS growth rates that
simplify the~computations of the growth rates for high energy beams.

\subsection{High energy approximations}
The complexities of the formulae developed by Piwinski and by Bjorken and Mtingwa for
the IBS growth rates have motivated efforts to find simplified expressions, to allow faster
and more convenient computation of the growth rates.  Computational efficiency is
important when calculating how the emittance in a storage ring changes over
time: the IBS growth rates depend on the beam emittance, so as the emittance changes,
the growth rates must be recalculated. In electron storage rings the equilibrium emittance
is determined by the balance between synchrotron radiation damping and emittance growth
from quantum excitation and IBS; since the IBS growth rates depend on the emittance,
some iteration is needed to find the value of the emittance for which all the effects balance.
Computational efficiency is again a~consideration in performing the calculations.

An important step in reducing the time needed to calculate the IBS growth rates is the replacement
of the integral in Eq.~(\ref{ibspiwinskiffunction}) that appears in Piwinski's formulae with a function that
can be more rapidly evaluated.  There are various approximations that may be made to this
function; here, we mention the~high-energy approximation derived by Bane \cite{bane2002}
and a later high-energy approximation derived by Mtingwa et al.~\cite{cimp}.

Bane started from the Bjorken--Mtingwa formulae to find formulae for the IBS
growth rates that can be written as follows:
\begin{eqnarray}
\frac{1}{T_z} & \approx & \frac{r_0^2 c N \mathrm{(log)}}{16 \gamma^3 \varepsilon_x^{3/4} \varepsilon_y^{3/4}
\sigma_z \sigma_\delta^3}
\langle \sigma_H g_\mathrm{Bane}(a/b) (\beta_x \beta_y)^{-1/4} \rangle, \\
\frac{1}{T_{x,y}} & \approx & \frac{\sigma_\delta^2 \langle \mathcal{H}_{x,y} \rangle}{\varepsilon_{x,y}}
\frac{1}{T_z},
\end{eqnarray}
where:
\begin{equation}
a = \frac{\sigma_H}{\gamma} \sqrt{\frac{\beta_x}{\varepsilon_x}}, \qquad
b = \frac{\sigma_H}{\gamma} \sqrt{\frac{\beta_y}{\varepsilon_y}}, \qquad
\frac{1}{\sigma_H^2} = \frac{1}{\sigma_\delta^2}
 + \frac{\mathcal{H}_x}{\varepsilon_x} + \frac{\mathcal{H}_y}{\varepsilon_y},
\end{equation}
and the function $g_\mathrm{Bane}$ is given by:
\begin{equation}
g_\mathrm{Bane}(\alpha) = \frac{2\sqrt{\alpha}}{\pi} \int_0^\infty
\frac{du}{\sqrt{1 + u^2}\sqrt{\alpha^2 + u^2}}.
\end{equation}
A numerical fit to $g_\mathrm{Bane}(\alpha)$ leads to the approximation:
\begin{equation}
g_\mathrm{Bane}(\alpha) \approx \alpha^{(0.021 - 0.044 \ln(\alpha))}.
\end{equation}
Bane's formulae are valid for $a$, $b\ll 1$ (which will be the case if the beam
is cooler longitudinally than transversely), and if effects from the gradients
of the lattice functions (dispersion and beta functions) are negligible.

\begin{figure}
\begin{center}
\includegraphics[width=0.85\textwidth]{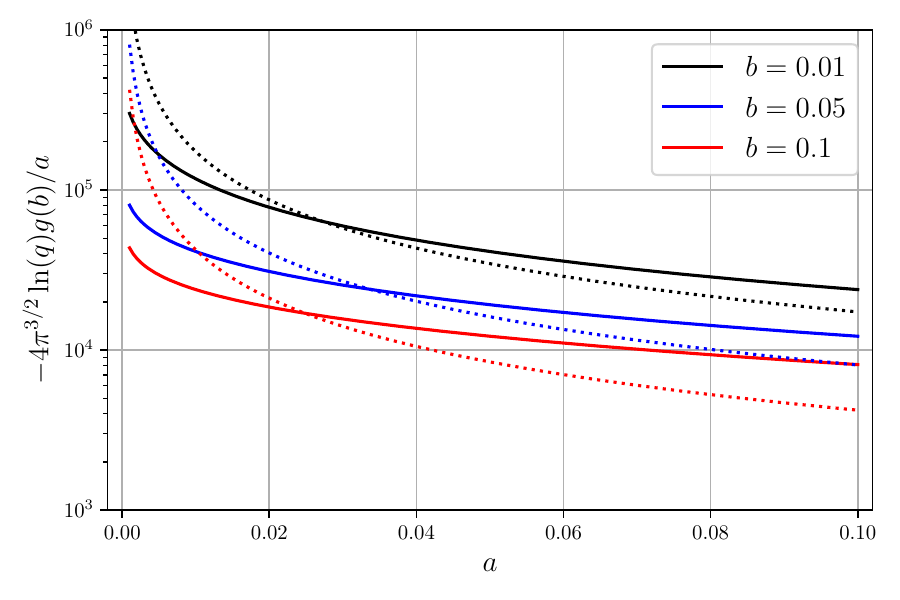}
\caption{Piwinski function $f$ (solid lines) compared with the high-energy
approximation (\ref{cimpf}) (dotted lines).  In this case, $q = 10^3$.
The Piwinski function is shown in more detail (and over a larger range of
parameters) in Fig.~\ref{figpiwinskif}.
\label{figpiwinskifapprox}}
\end{center}
\end{figure}

Alternative expressions for the (approximate) IBS growth rates, again avoiding
the integral in Piwinski's formulae, were obtained by Mtingwa et al.~\cite{cimp}
using an earlier approximation for the function $f(a,b,q)$ by Mtingwa and Tollestrop
\cite{mtingwatollestrup1987}:
\begin{equation}
f(a,b,q) \approx -\frac{4\pi^{3/2}}{a}\,g(b)\,\ln(q),
\label{cimpf}
\end{equation}
where $g(b)$ is given by:
\begin{equation}
g(b) = \sqrt{\frac{\pi}{b}} \left(
P^0_{-1/2}(\theta) + \sgn(b-1) \frac{3}{2} P^{-1}_{-1/2}(\theta)
\right).
\end{equation}
In this expression, $\theta = (b^2 + 1)/2b$, and $P^\mu_\nu(\theta)$ are associated
Legendre functions. Unfortunately, the~functions $P^\mu_\nu(\theta)$ are not easy
to evaluate, and this can further limit the advantages of the approximation (see Eq.~(\ref{cimpf}))
compared to the use of the full expression (see Eq.~(\ref{ibspiwinskiffunction})) involving an integral.
However, use of Eq.~(\ref{cimpf}) leads to the ``completely integrated modified Piwinski''
(CIMP) formulae for the IBS growth rates:
\begin{eqnarray}
\frac{1}{\tau_x} & \approx & 2\pi^{\frac{3}{2}}A \left\langle
-a \ln\!\left( \frac{q^2}{a^2} \right) g\!\left( \frac{b}{a} \right) +
\frac{\mathcal{H}_x\sigma_H^2}{\epsilon_x}
\, G(a,b,q)
\right\rangle \\
\frac{1}{\tau_y} & \approx & 2\pi^{\frac{3}{2}}A \left\langle
-b \ln\!\left( \frac{q^2}{b^2} \right) g\!\left( \frac{a}{b} \right) +
\frac{\mathcal{H}_y\sigma_H^2}{\epsilon_y}
\, G(a,b,q)
\right\rangle \\
\frac{1}{\tau_z} & \approx & 2\pi^{\frac{3}{2}}A \left\langle
\frac{\sigma_H^2}{\sigma_p^2} \, G(a,b,q)
\right\rangle .
\end{eqnarray}
The function $G(a,b,q)$ is given by:
\begin{equation}
G(a,b,q) = \frac{\ln\!\left( \frac{q^2}{a^2} \right) g\!\left( \frac{b}{a} \right)}{a} +
\frac{\ln\!\left( \frac{q^2}{b^2} \right) g\!\left( \frac{a}{b} \right)}{b},
\end{equation}
with:
\begin{equation}
a = \frac{\sigma_H}{\gamma_0} \sqrt{\frac{\beta_x}{\epsilon_x}}, \quad 
b = \frac{\sigma_H}{\gamma_0} \sqrt{\frac{\beta_y}{\epsilon_y}}, \quad 
q = \beta_0 \sigma_H \sqrt{\frac{2\beta_y\epsilon_y}{r_0}}, \label{ibspiwinskiq1mod}
\end{equation}
and:
\begin{equation}
\frac{1}{\sigma_H^2} = \frac{1}{\sigma_\delta^2} + \frac{\mathcal{H}_x}{\epsilon_x} + \frac{\mathcal{H}_y}{\epsilon_y}.
\label{ibspiwinskisigmah1mod}
\end{equation}
The approximations leading to these results are again valid in the high energy
regime, where $a$ and $b$ are much smaller than $q$.  A comparison between
the Piwinski function $f(a,b,q)$ and the approximation (\ref{cimpf}) is shown
in Fig.~\ref{figpiwinskifapprox}.  For a numerical comparison of the IBS growth
rates calculated using the full Piwinski formulae, Bane's approximation and the
CIMP approximation in a particular case, see \cite{cimp}.

\newpage
\section{Tests of IBS theory against experimental measurements\label{sec:experimentalmeasurements}}

Although the effects of intrabeam scattering have long been observed and there has been
extensive work on understanding the theory and on developing mathematical models, validation of the
formulae for the emittance growth rates has been difficult.  There are two main reasons for this. 
First, although emittance growth from IBS can be significant, the effects of IBS are often
weak relative to other phenomena that may lead to changes in emittance (such as
space charge effects, nonlinear behaviour resulting from higher-order multipole components
in the magnets, or wakefield effects).  Separating IBS emittance growth from growth caused by
other effects is not straightforward.  Secondly, the IBS emittance growth depends on
many beam and lattice properties: without a full and detailed knowledge of all the relevant
quantities the IBS emittance growth can only be calculated from the theory by making
some assumptions, and this inevitably leads to some uncertainty in the results.

Despite the challenges in testing the theoretical models of IBS,
quantitative studies have been carried out over the years in a variety of machines, covering
a wide range of parameter regimes.  Here, we mention briefly a few of those studies.

\subsection{KEK-ATF}

\begin{figure}
\begin{center}
\includegraphics[width=0.65\textwidth]{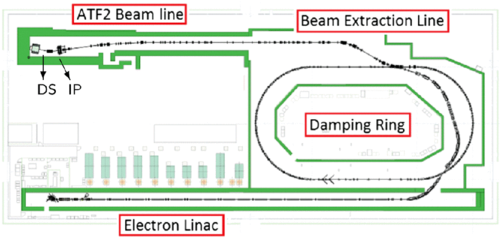} \\[0.2in]
\includegraphics[width=0.65\textwidth]{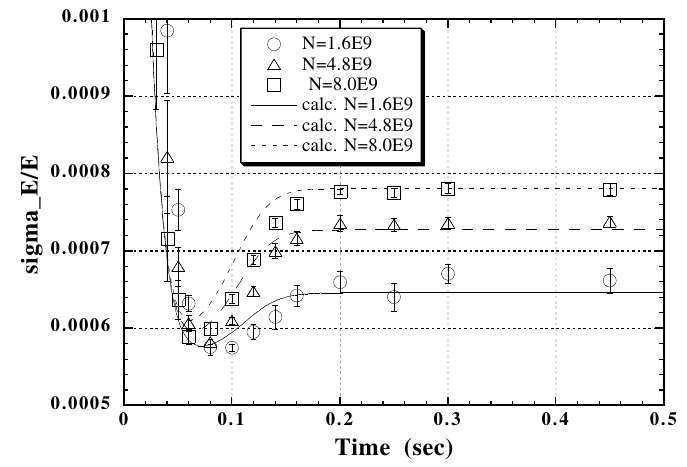}
\caption{Top: layout of the KEK Accelerator Test Facility (ATF), showing the 1.28\,GeV electron
linac, the storage (or damping) ring, and the extraction line.  Bunches can be extracted from
the ring at any time following injection; high dispersion in the extraction line allow accurate
measurements of beam energy spread to be made. Bottom:~evolution of the
energy spread of bunches following injection into the ATF.  Three different bunch populations
($1.6\times 10^9$ for circular markers, $4.8\times 10^9$ for trangular markers and
$8.0\times 10^9$ for square markers) are shown.  Markers show experimental data; lines
show theoretical expectations, taking into account synchrotron radiation damping and
intrabeam scattering.  Because the longitudinal damping time is shorter than the transverse
damping times, the~energy spread reaches a very low value while the vertical emittance is
still relatively large; as the vertical emittance continues to damp, the IBS growth rates increase
leading to an increase in the energy spread.
\label{atfibsexperiment}}
\end{center}
\end{figure}

The Accelerator Test Facility (ATF) at KEK was developed to support research for damping
rings for a~future linear collider.  It operates as a low-emittance storage ring with full-energy
injection at 1.28\,GeV.  A suite of diagnostics provide detailed information about the beam
properties, including energy spread and emittance.  A particular goal was the demonstration
of vertical emittance of a few picometres.  In 2002, a detailed study was carried out
to characterise the damping of the beam emittances after injection.  It was found that initially
the emittances reduced at the rate expected from synchrotron radiation effects, but that
the longitudinal emittance, rather than damping monotonically to an equilibrium value
(as would be expected from radiation damping),
passed through a minimum about 100\,ms after beam injection, before increasing to a final
equilibrium that depended on the bunch charge.  This behaviour can be seen in the results
shown in Fig.~\ref{atfibsexperiment}.

The fact that the longitudinal emittance passes through a minimum following injection (before
reaching some equilibrium value above the minimum) can be understood by considering the
effects of IBS.  In an electron storage ring,
the radiation damping time in the longitudinal direction is usually about half the transverse
damping time: this means that the longitudinal emittance can approach an equilibrium value
much sooner than the transverse emittances.  In the case of the ATF, IBS effects are negligible
in an injected bunch with large emittances, and become significant only when the transverse
emittances reach low values.  A vertical emittance of a few picometres leads to IBS emittance
growth rates that are comparable with the radiation damping rates.  As a result, once the
vertical emittance approaches its equilibrium value (between 50\,ms and 100\,ms after injection
into the ATF storage ring) IBS effects become strong enough to increase the longitudinal
emittance.  The final longitudinal and transverse emittance values depend on the bunch charge,
which is readily controlled in the ATF.

The plot in Fig.~\ref{atfibsexperiment} \cite{kubo2002} shows a comparison between
measurements (points
with circular, triangular or square markers on the plot, according to the bunch charge) and
calculations based on the Bjorken--Mtingwa formalism \cite{sadibs}.  Although the agreement
is not perfect, the behaviour of the longitudinal emittance is described well using the theory.

\subsection{CELSIUS\label{subsectioncelsius}}

The CELSIUS storage ring \cite{ekstrom1988} operated between 1989 and 2005
at the Svedberg Laboratory, Uppsala University, and was part of an accelerator
facility for research into nuclear physics.  The storage ring was designed to cool
and accelerate beams covering a range of species and energies, from protons
(up to 1360\,MeV kinetic energy) to ions with $A \sim 100$ (470\,MeV per nucleon
for particles with charge to mass ratio 1/2).  Especially at the lower end of the
energy range, intrabeam scattering effects could be significant.  In normal operation,
the beam emittances were damped by electron cooling; but the ability to turn off
the cooling provided the ability to make measurements of IBS emittance growth in
a range of different parameter regimes.

\begin{figure}
\begin{center}
\includegraphics[width=0.32\textwidth]{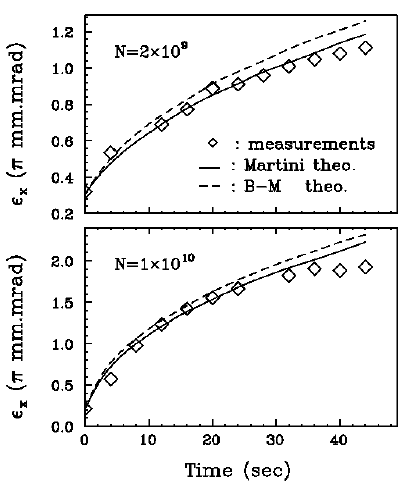}
\includegraphics[width=0.32\textwidth]{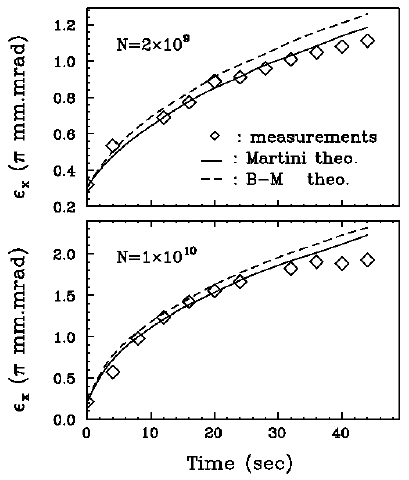}
\includegraphics[width=0.32\textwidth]{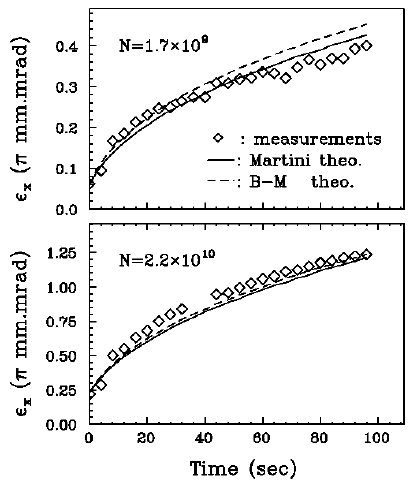}
\caption{Measurements of intrabeam scattering in the CELSIUS storage ring \cite{raohermansson2000}.
Each plot shows the horizontal emittance as a function of time after turning off the transverse
beam cooling. Measurements (diamond-shaped markers) are compared with calculations
based on the formulae of Martini (solid lines) and Bjorken--Mtingwa (dashed lines).
Left: 48\,MeV proton beam with bunch populations $2\times 10^9$ (top) and
$1\times 10^{10}$ (bottom).
Middle: 400\,MeV proton beam with bunch populations $1.7\times 10^9$ (top) and
$2.2\times 10^{10}$ (bottom).
Right: 200\,MeV/u beam of N$^{7+}$ ions with bunch populations $8.3\times 10^8$ (top) and
$1.7\times 10^9$ (bottom).
\label{celsiusexperiment2000}}
\end{center}
\end{figure}

Some results from emittance measurements at CELSIUS are shown in Fig.~\ref{celsiusexperiment2000} \cite{raohermansson2000}.  Each plot shows a comparison
between measurements of the horizontal emittance as a function of time after switching
off the beam cooling and the predictions from IBS theory under the corresponding conditions.
Results are shown for 48\,MeV protons, 400\,MeV protons, and 200\,MeV/u N$^{7+}$
ions, with two different bunch populations in each case.  Measurements were made with
the storage ring operating above the transition energy. Theoretical predictions, made using
Martini's formulae and on the formulae of Bjorken and Mtingwa, are in good agreement
with the measurements.  It should be noted, however,
that only the horizontal emittance was measured: for the calculations, the bunch length
and energy spread were assumed to be equal to nominal values based on values measured
for injected bunches and reduced according to the~expected effects of the beam cooling.
The vertical emittance was assumed to be equal to the horizontal emittance.  For further
details, the reader is referred to \cite{raohermansson2000}.

\subsection{Gold Ions in RHIC}
The measurements of IBS emittance growth in CELSIUS shown in Section~\ref{subsectioncelsius}
were made with the~storage ring operating above transition.  Measurements of the effects of IBS on
the emittances of beams in a storage ring below transition have also been made, for example
in the Relativistic Heavy Ion Collider (RHIC) at Brookhaven \cite{fischer2001}.  IBS was of
particular concern in RHIC because of the fact that ion beams were generally injected at
low energy, below transition in the storage ring, before the energy was ramped to the full
energy required for collider operation; acceleration involved crossing the transition energy
(with associated manipulations of the longitudinal phase space) and any increase in longitudinal
emittance of the beams after injection could lead to particle loss at transition crossing.

\begin{figure}
\begin{center}
\includegraphics[width=0.47\textwidth]{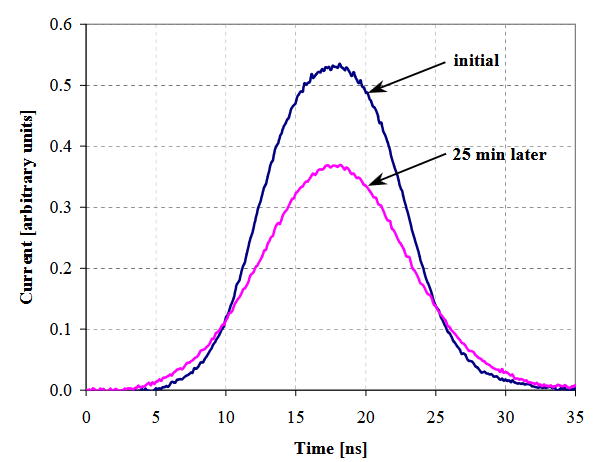}
\includegraphics[width=0.47\textwidth]{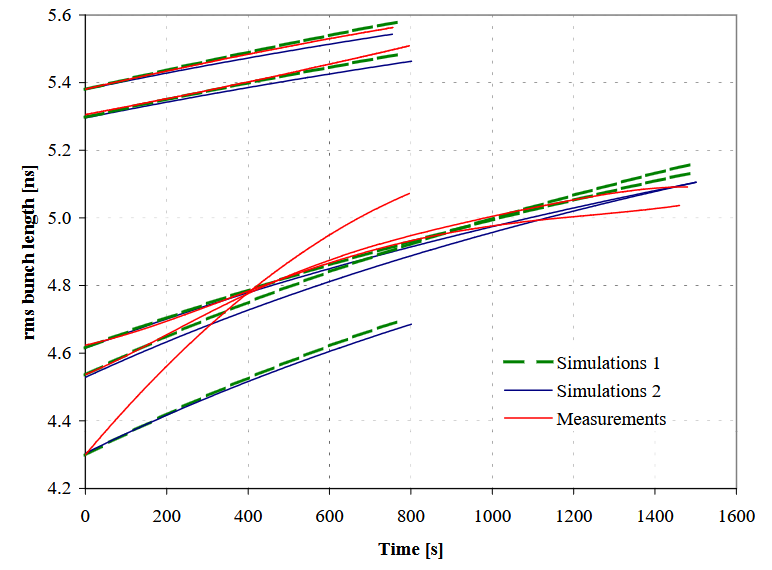}
\caption{Measurements of intrabeam scattering in beams of gold ions in RHIC 
at injection energy (relativistic $\gamma = 10.25$) \cite{fischer2001}.  At injection,
the ring is operating below transition.  Left: longitudinal charge profile immediately
after injection (blue line) and after 25 minutes (magenta line). Right: increase of
bunch length with time for five different sets of measurements.  Results from two
different simulation codes (dashed green line and solid blue line) are compared with
measurements (solid red line).
\label{rhicexperiment2001}}
\end{center}
\end{figure}

In 2000, a set of experiments was performed using gold ions. Measurements of bunch length and
transverse (vertical) beam size were made over a period of half an hour following beam injection,
with the ions having a relativistic factor $\gamma = 10.25$.  Bunch length measurements were made
using a wall current monitor, which gave the charge density as a function of time with a resolution
of 0.25\,ns; typical results are shown in Fig.~\ref{rhicexperiment2001}, left.  Transverse beam
size measurements were made using ionization beam profile monitors \cite{cameron1999}.
Because of technical issues at the time of the measurements, only the vertical beam size
was measured; to compare the results of measurements with simulations, it was assumed that the~horizontal beam size was equal to the vertical beam size.  Simulations were performed with two
different computer codes, both making high-energy approximations to the IBS growth rate formulae
\cite{wei1993, parzen1987}. A~comparison between measurements and simulations of the growth in
the bunch length is shown for five different cases in Fig.~\ref{rhicexperiment2001}, right.

In most cases, there is reasonable agreement between the measured growth rates and the
growth rates predicted by the computer codes.  In the case showing a significant difference
(with a measured growth rate about twice the growth rate predicted by the computer codes)
it is thought that some mechanism other than intrabeam scattering may have contributed to
the growth in the bunch length.

In a ring operating below transition, the theory predicts that intrabeam scattering should lead
to a~redistribution of the emittances, with the sum of the emittances remaining bounded
according to Eq.~(\ref{conservationofenergy3}).  Therefore, in the case of the measurements
made in RHIC, with an increasing bunch length it is expected that there should be a fall in the
transverse emittance.  However, only one of the computer codes predicted a reduction in
transverse emittance; for reasons that are not completely clear (but that may be to do with
the use of a high energy approximation in the calculations) the other code predicted a slow
growth in transverse emittance.  The measurements also showed an increase in transverse
emittance, though it should be remembered that only the vertical emittance was measured during
the experiments.  The reason for this behaviour is again not clear, though it is again possible
that effects other than intrabeam scattering were contributing to the emittance growth.

\subsection{Protons in the Fermilab Recycler ring}
\begin{figure}
\begin{center}
\includegraphics[width=0.48\textwidth]{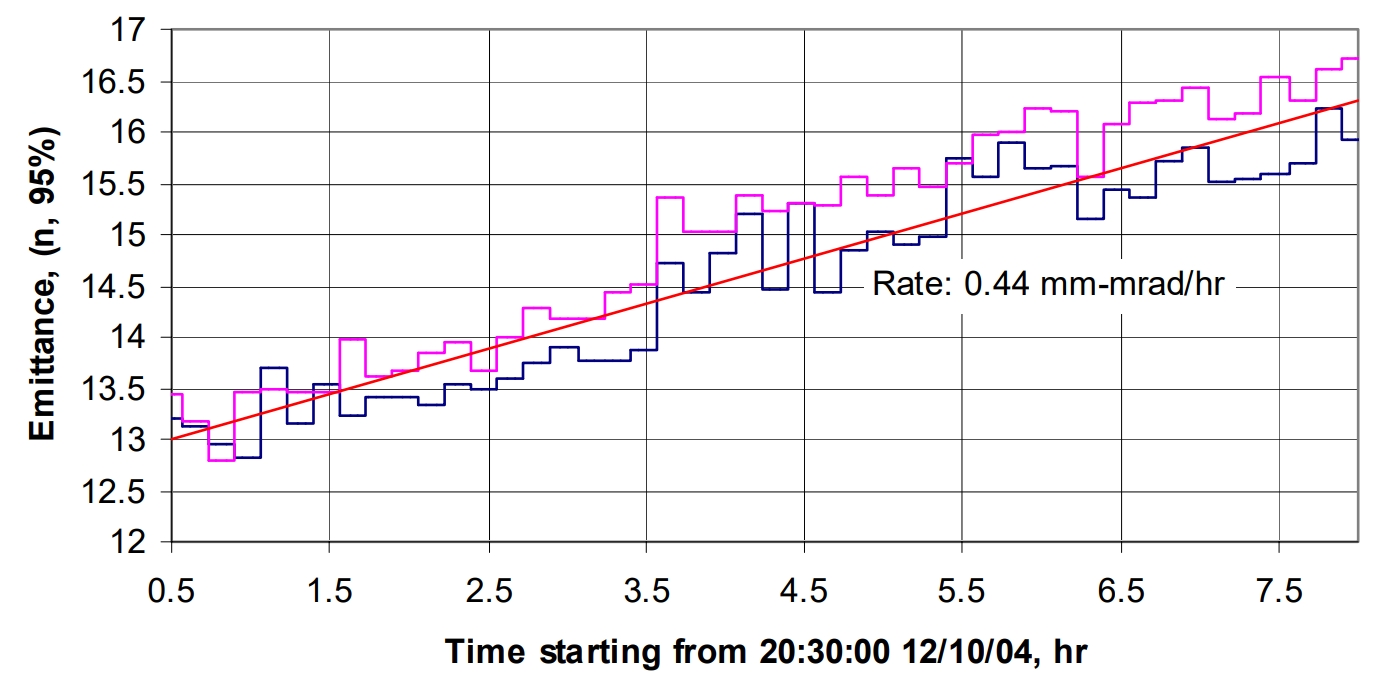}
\includegraphics[width=0.47\textwidth]{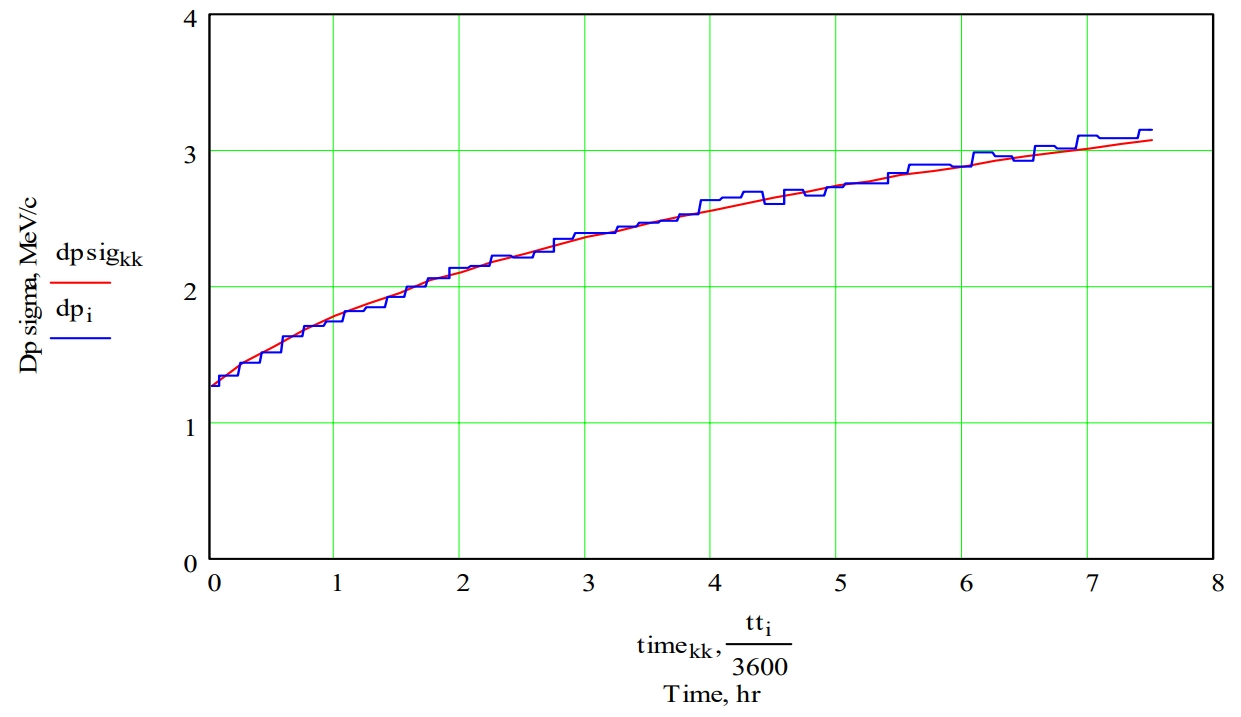} \\
(a) Proton bunch population $100\times 10^{10}$. \\[0.2in]
\includegraphics[width=0.47\textwidth]{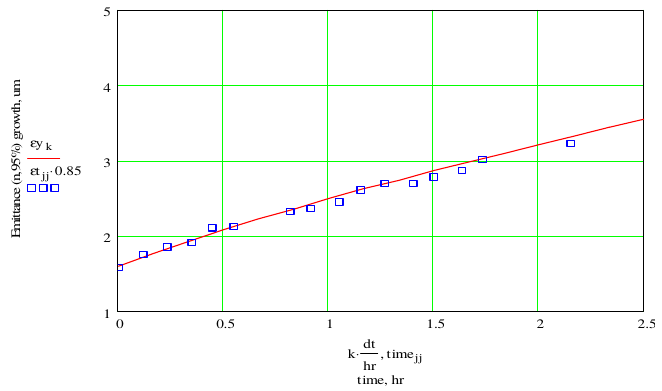}
\includegraphics[width=0.47\textwidth]{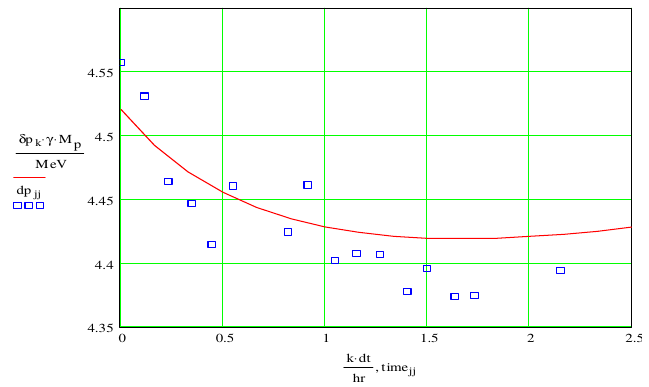} \\
(b) Antiproton bunch population $25\times 10^{10}$.
\caption{Measurements of intrabeam scattering in the Fermilab Recycler at 9.8\,GeV/c beam
momentum (below transition) \cite{hu2005}.  Top: transverse emittance (left) and momentum
spread (right) as functions of time in a bunch of $100\times 10^{10}$ protons.  With an initially
low energy spread, IBS leads to an increase in longitudinal emittance.  The
transverse emittance also increases, but at a slower rate than would be expected from gas
scattering under the relevant conditions.  Bottom: transverse emittance (left) and momentum
spread (right) as functions of time in a~bunch of $25\times 10^{10}$ antiprotons.  In this case,
the initial longitudinal emittance (right) is relatively large, and reduces over time from the effects of
IBS. The growth rate of the transverse emittance (left) is increased in this case above the rate
expected from gas scattering.  
\label{fermilabrecycler2005}}
\end{center}
\end{figure}

Measurements of IBS emittance growth in a storage ring below transition were also made
using the~Fermilab Recycler, a proton/antiproton storage ring operating at a fixed beam
momentum of 9.8\,GeV/c.  The~beam momentum at transition is 19.4\,GeV/c.  In preparation
for the Run II luminosity upgrade of the Tevatron, experiments were carried out to
characterise the effects of IBS on the longitudinal and transverse emittances of beams in
the Recycler ring: results are reported in Ref.~\cite{hu2005}.  The momentum spread was of
particular interest because of the need to produce beams with a small longitudinal
emittance in the~Tevatron.  Measurements in the Recycler were first performed with a very low
initial momentum spread (leading to longitudinal heating) and then with a larger initial
momentum spread (resulting in longitudinal cooling).  Some results are shown in
Fig.~\ref{fermilabrecycler2005}.

The upper two plots in Fig.~\ref{fermilabrecycler2005} show the results of measurements
with a bunch of $100\times 10^{10}$~protons, with transverse emittance measurements in
the left-hand plot, and energy spread measurements on the right.  In this case, the initial
energy spread is small and IBS leads to an increase in the longitudinal emittance.  With the
ring below transition, it is then expected that there will some cooling in the transverse
emittance.  The measurements actually show a growth over time in the transverse
emittance, at a~rate of 0.44$\pi$\,mm\,mrad/hr; but this is a lower rate of growth than is
measured under similar conditions but with a bunch population of $1\times 10^{10}$.
With the lower bunch charge, a transverse emittance growth rate of 0.60$\pi$\,mm\,mrad/hr
is observed: with this bunch population effects of intrabeam scattering are negligible, and
the transverse emittance growth is the result of scattering between particles in the bunch
and residual gas in the vacuum chamber.  The emittance growth from gas scattering is
independent of bunch charge: intrabeam scattering in this case, therefore, leads to a
cooling of the transverse emittance as expected from the theory.

The lower two plots in Fig.~\ref{fermilabrecycler2005} show the results of measurements
with a bunch of $25\times 10^{10}$ antiprotons, with transverse emittance in the left-hand
plot and momentum spread in the right-hand plot.  In this case, the initial momentum
spread is much larger than in the previous case ($100\times 10^{10}$ protons per bunch),
and IBS leads to longitudinal cooling.  The red line in the left hand plot in
Fig.~\ref{fermilabrecycler2005} (b) shows the~predicted emittance growth from IBS and
gas scattering. The growth rate from gas scattering is the only free parameter in the fit: the
best fit is obtained with a gas scattering growth rate of 0.60$\pi$\,mm\,mrad/hr, which is
consistent with separate measurements (at low bunch population).  With IBS, the growth
rate is close to 0.9$\pi$\,mm\,mrad/hr, so IBS makes a positive contribution to the transverse
emittance growth.  Again, this is consistent with the theory which predicts transverse
heating under conditions where there is longitudinal cooling.

\subsection{The FERMI Free-Electron Laser}
Observations of IBS effects have also been made in the driver for the~FERMI free-electron laser
\cite{dimitri2020}, in a very different parameter regime from those in the machines mentioned
so far.  Like the KEK-ATF, the~FERMI FEL is an electron accelerator; but whereas the KEK-ATF is a
storage ring the FERMI FEL is a single-pass beamline.  With improvements in technology in
recent years, FEL injector systems have become capable of producing electron bunches with charge densities high
enough for IBS effects to be significant on the short timescales associated with single-pass machines.

For a short-wavelength (VUV or X-ray) FEL to operate effectively, it is important to achieve a~short
bunch length and  low energy spread in the electron bunch reaching the undulator.  Considerable
efforts are made to understand and control both the bunch length and the energy spread, and
detailed models including all relevant physical effects are needed if an accurate match between
simulations and measurements is to be achieved.  The FERMI FEL, like other short-wavelength
free-electron lasers, relies on bunch compressors to reduce the bunch length to the range
required for FEL operation.  Bunch compressors work by rotating the longitudinal phase space,
and necessarily lead to an increase in energy spread: in the absence of collective effects,
the energy spread increases by the same factor by which the~bunch length is reduced, so that
the longitudinal emittance (essentially, the product of the bunch length and the energy spread)
is conserved.  However, with high charge and short bunches, emission of coherent synchrotron
radiation (CSR) in the dipoles in a bunch compressor can act back on particles in the bunch, leading
to an increase in the energy spread far beyond the limit needed for FEL operation.  The~effect of
CSR may be reduced by Landau damping: artificially increasing the energy spread (e.g.~by
interacting the bunch with a laser pulse in a `laser heater') disrupts the coherent motion of the
particles and suppresses emission of CSR.  Although use of a laser heater leads to some increase
in the energy spread, the increase can be significantly less than that resulting from CSR in the
absence of a laser heater.

\begin{figure}
\begin{center}
\includegraphics[width=0.32\textwidth]{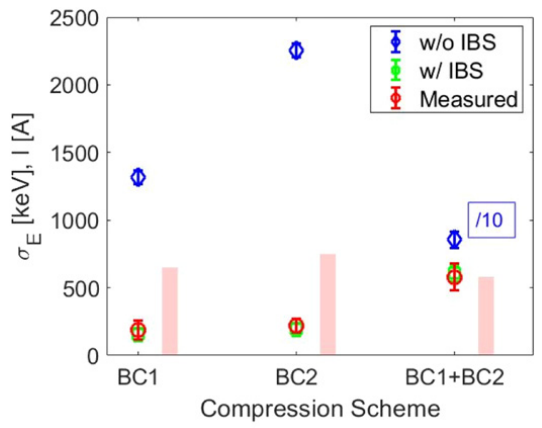}
\includegraphics[width=0.32\textwidth]{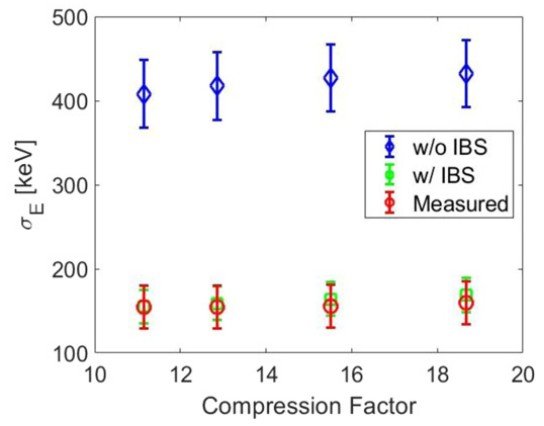}
\includegraphics[width=0.30\textwidth]{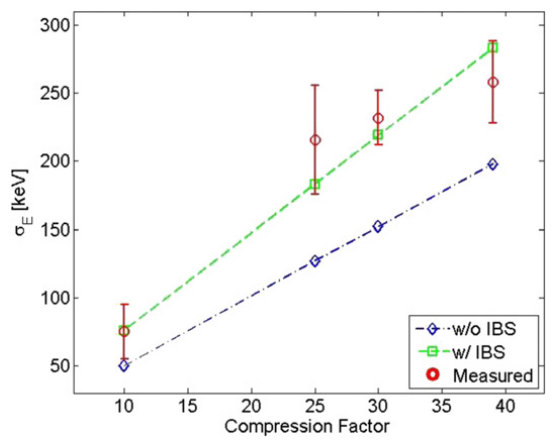}
\caption{Observations of intrabeam scattering in the FERMI FEL driver with
electron bunches accelerated to energy between 713\,MeV and 754\,MeV \cite{dimitri2020}.
Left: without use of a laser heater to suppress coherent synchrotron radiation (CSR) by increasing
the energy spread, the measured energy spread (red markers) with each of three bunch
compression schemes is significantly below the energy spread that would be expected in
the absence of IBS (blue markers).  Including IBS in the simulations leads to energy spreads
in good agreement with measurement: the increase in energy spread at low energy from
IBS suppresses the effects of CSR that can lead to much larger increases in energy spread.
Middle: energy spread as a function of compression factor in BC1, with bunch charge 650\,pC
and with laser heater turned off.  The energy spread is dominated by CSR so does not
increase with the~compression factor; but the impact of CSR is reduced by the increase in
energy spread caused by IBS.
Right: energy spread as a function of compression factor in BC1, with bunch charge 100\,pC
and with laser heater turned on.  In this case, the laser heater almost entirely suppresses the
effects of CSR so the energy spread increases with compression factor as expected; an increase
in energy spread from IBS can now also be observed.
\label{fermifel2020}}
\end{center}
\end{figure}

Figure \ref{fermifel2020} shows a comparison between energy spread measurements at the end
of the FERMI linac (beam energy between 713 and 754\,MeV with 100\,pC bunch charge,
or 900\,MeV with 650\,pC bunch charge) and the results of simulations.  The left-hand plot shows
the case of a 100\,pC bunch, with laser heater turned off; results for three different compression
schemes are shown.  Without IBS (blue points) simulations predict energy spread above 1200\,keV:
this is in large part due to CSR.  Note that for the case of the combined bunch compression scheme
(BC1+BC2) the point shown on the plot has been reduced by a factor of 10 to fit on the same axis
as the other cases.  If IBS effects are included, there is an effective suppression of CSR, and the
results of the simulations are in good agreement with the~measurements.

The central plot in Fig.~\ref{fermifel2020} shows results for the case of a 650\,pC bunch, again
with the laser heater turned off.  In this case, only the first bunch compressor (BC1) is used, but
with a range of different compression factors.  Without IBS, CSR is again expected to make a
dominant contribution to the final energy spread; including IBS in the simulations shows that
the consequent modest increase in energy spread is sufficient to suppress CSR to a significant
degree.  The results of simulations including IBS are in good agreement with the energy spread
measurements.

Finally, the right-hand plot in Fig.~\ref{fermifel2020} shows the energy spread in the case of a
100\,pC bunch with the laser heater turned on, for different compression factors in BC1.  The
laser heater suppresses CSR almost entirely: in this case, we see the expected increase in energy
spread with increasing compression factor.  Under these conditions, IBS no longer has a beneficial
effect in suppressing CSR (which has already been largely eliminated by the laser heater) but
instead makes a positive contribution to the energy spread.  The measured energy spread is larger
than expected from simulations in which IBS is omitted; however, although the agreement is not perfect, the
simulations are consistent with the measurement results if IBS is included.

\section{Final remarks}
Intrabeam scattering has continued to be much studied since the initial observations and
theoretical investigations in the 1960's and 1970's.  Even though the fundamental processes are
well understood, predicting the effects of IBS in specific cases remains difficult
because of the complexity of the calculations.  Experimental studies are challenging
because of the difficulty in making precise measurements of all the~relevant parameters, and
also because other effects can lead to changes in emittance on the timescale of the changes
that would be expected from IBS.

Historically, IBS has been of greater significance in hadron accelerators than in electron machines.
This is because IBS effects have generally been too slow to have much impact in single-pass
systems; and in storage rings, radiation damping in the case of electrons dominates over emittance
growth from IBS.  In recent years, however, electron storage rings have started to operate in
regimes where emittance growth rates from IBS have an impact on the equilibrium beam
properties -- this is particularly the case for the latest generation of synchrotron light sources,
operating with ultra-low emittances (of the order of tens of picometres).  IBS effects have now
even been observed in electron beams in a single-pass system, namely the driver for the FERMI
FEL.  Meanwhile, experimental studies in hadron storage rings have been carried out over a wide
range of conditions, including in proton, antiproton and ion beams, and in rings operating above
and below transition.  The overall picture that emerges is one that generally supports the validity
of the IBS theories.

The first detailed theoretical treatment of IBS, by Piwinski in the early 1970's, continues to be
widely used as the basis for computing emittance growth rates from IBS in many different
parameter regimes, albeit with some extensions to include aspects omitted in the original
treatment (such as the variation of beta functions around the circumference of a storage ring).
The theory of Bjorken and Mtingwa, published in the early 1980's, provides an alternative
formalism.  Other researchers have contributed in attempting to provide approximations that
can be used in certain regimes (notably, at high energy) to simplify the calculations and speed
up the computations.  Despite dramatic improvements in computing
power over the years, work to develop formulae allowing rapid, accurate computation of IBS
growth rates seems likely to continue.

Understanding IBS has been important for the design and operation of many different
accelerators.  The push to parameter regimes with ever-higher charge density (high bunch
charge with low emittance and bunch length) seems likely to make IBS of increasing
significance in future facilities, including hadron and electron storage rings, and even in
single-pass machines such as free-electron lasers.

\end{document}